\documentclass[sigplan,twocolumn,10pt]{acmart}
\AtBeginDocument{%
  }

\usepackage{graphicx}
\usepackage{float}  
\usepackage{subcaption}  
\usepackage[ruled,vlined,linesnumbered]{algorithm2e}

\usepackage{hyperref}
\usepackage{xspace} 
\usepackage{amsmath}

\usepackage{enumitem} 

\usepackage{pifont}

\acmSubmissionID{257}

\newcommand{\eqnref}[1]{\hyperref[#1]{Eq.~(\ref{#1})}}  
\newcommand{\figref}[1]{\hyperref[#1]{Fig.~\ref*{#1}}}
\newcommand{\tabref}[1]{\hyperref[#1]{Tab.~\ref*{#1}}}
\newcommand{\algref}[1]{\hyperref[#1]{Alg.~\ref*{#1}}}

\makeatletter
\renewcommand*{\p@section}{\S} 
\renewcommand*{\p@subsection}{\S}
\renewcommand*{\p@subsubsection}{\S}

\makeatother

\usepackage{tikz} 
\newcommand*\circled[1]{
  \tikz[baseline=(char.base)]{
    \node[shape=circle,draw,thick,inner sep=0pt] (char) {\sffamily\bfseries \footnotesize #1};}}

\newcommand{\circledTitle}[2]{
  \noindent\textbf{\circled{\footnotesize #1}: \textit{ #2}}
}

\newcommand{\parab}[1]{\noindent\textbf{#1}}

\begin{document}

\newcommand{\name}{\textsf{LCMP}\xspace}
\newcommand{\reviseadd}[2]{{\color{red} [\textbf{#1}] #2\xspace}}
\renewcommand{\reviseadd}[2]{#2}

\title[LCMP]{LCMP: Distributed Long-Haul Cost-Aware Multi-Path Routing for Inter-Datacenter RDMA Networks}

\author{Dong-Yang Yu}
\orcid{0009-0003-3830-9388}
\additionalaffiliation{
\institution{State Key Laboratory of Networking and Switching Technology}
}
\affiliation{
\institution{BUPT}
\country{China}
}

\author{Yuchao Zhang}
\orcid{0000-0002-0135-8915}
\authornotemark[1]
\authornote{Yuchao Zhang is the corresponding author.}
\affiliation{
\institution{BUPT}
\country{China}
}

\author{Xiaodi Wang}
\orcid{0009-0004-1035-9559}
\affiliation{
\institution{BUPT}
\country{China}
}
\authornotemark[1]

\author{Jun Wang}
\orcid{0009-0001-0287-4399}
\affiliation{
\institution{BUPT}
\country{China}
}
\authornotemark[1]

\author{Wenfei Wu}
\orcid{0000-0002-1357-3137}
\affiliation{%
  \institution{Peking University}
  \country{China}
}

\author{Haipeng Yao}
\orcid{0000-0003-1391-7363}
\affiliation{%
  \institution{BUPT}
  \country{China}
}

\author{Wendong Wang}
\orcid{0000-0002-6418-8087}
\affiliation{
\institution{BUPT}
\country{China}
}
\authornotemark[1]

\author{Ke Xu}
\orcid{0000-0003-2587-8517}
\affiliation{%
  \institution{Tsinghua University}
  \country{China}
}
\affiliation{%
  \institution{Zhongguancun Laboratory}
  \country{China}
}


\renewcommand{\shortauthors}{Dong-Yang Yu et al.}

\newcommand{\ydy}[1]{{\color{red} #1\xspace}}
\renewcommand{\ydy}[1]{#1}

\begin{abstract}
RDMA-empowered cloud services are gradually deployed across datacenters (DCs) with multiple paths, which exhibit new properties of path asymmetry, delayed congestion signals, and simultaneous flow routing collisions, and further fail existing routing methods.

We present \name, a distributed long-haul cost-aware multi-path routing framework that aims to place RDMA flows on multiple inter-DC paths, achieving low-cost, low-latency, and congestion-responsive transmission. \name combines a control-plane path-quality score with compact on-switch congestion signals, where the former unifies quality assessment for asymmetric paths and the latter enables responsive reaction to path congestion. \name further resolves the simultaneous flow decision collision problem by filtering high-cost candidates, and performing a diversity-preserving hash inside the reduced set. On an 8-DC testbed, \name reduces median and tail FCT slowdown by up to 76\% and 64\%, respectively compared to state-of-the-art (SOTA) baselines. And large-scale NS-3 simulations under the 2000 km inter-DC scenario confirm similar improvements.

\end{abstract}



\acmYear{2026}\copyrightyear{2026}
\setcopyright{cc}
\setcctype[4.0]{by}
\acmConference[EUROSYS '26]{21st European Conference on Computer Systems}{April 27--30, 2026}{Edinburgh, Scotland Uk}
\acmBooktitle{21st European Conference on Computer Systems (EUROSYS '26), April 27--30, 2026, Edinburgh, Scotland Uk}
\acmDOI{10.1145/3767295.3803593}
\acmISBN{979-8-4007-2212-7/2026/04}
\maketitle

\section{Introduction}
\label{sec:intro}

\begin{figure*}[t]
  \centering
    \begin{subfigure}[T]{2.4143 in}
        \includegraphics[width=\textwidth]{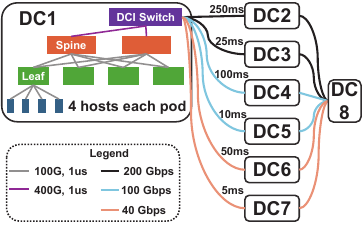}
        \caption{Inter 8-DC topology\protect\footnotemark.}
        \label{subfig:moti-porCor}
    \end{subfigure}
  \hfill
    \begin{subfigure}[T]{1.9034 in}
        \centering
        \includegraphics[width=\textwidth]{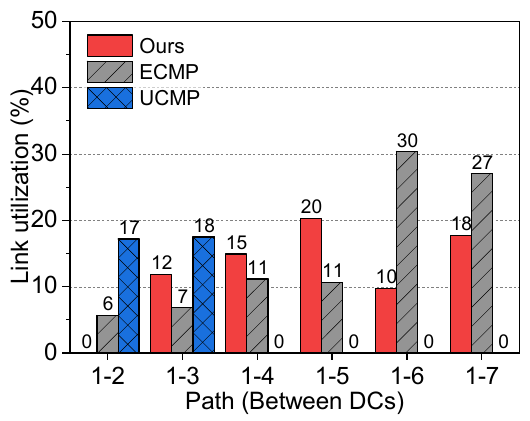}
        \caption{Per-link utilization.}
        \label{subfig:moti-util}
    \end{subfigure}
  \hfill
  \begin{subfigure}[T]{2.034 in}
    \centering
    \includegraphics[width=\textwidth]{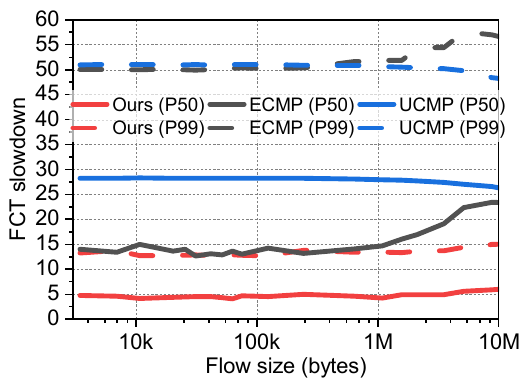}
    \caption{Median and tail FCT slowdown Median and tail FCT slowdown for Web Search under 30\% load using DCQCN.}
    \label{subfig:moti-FCT}
  \end{subfigure}
  \caption{[Motivation] Capacity–delay asymmetry causes ECMP and UCMP to make poor placement choices. \name balances utilization and reduces both median and tail FCT.}
  \Description{[Motivation] Capacity–delay asymmetry causes ECMP and UCMP to make poor placement choices; \name balances utilization and reduces both median and tail FCT.}
  \label{fig:moti-total}
\end{figure*}


\ydy{Modern cloud services increasingly depend on geographically distributed deployments that span multiple datacenters (DCs) to provide geo-replicated storage\cite{gao2021WhenCloud,bai2023EmpoweringAzure} and distributed machine learning training\cite{gangidi2024RDMAEthernet,bai2023EmpoweringAzure}, which impose stringent latency and throughput requirements while transferring large volumes of data across inter-DC links. To meet these demanding performance requirements, RDMA-empowered cloud services are being gradually deployed across DCs, leveraging RDMA's ability to offload the network stack to RNICs and bypass the kernel for ultra-low latency with minimal CPU overhead \cite{zhu2015CongestionControl, guo2016RDMACommodity}. However, as these RDMA flows traverse multiple inter-DC paths, they encounter new challenges including path asymmetry, outdated congestion signals, and simultaneous flow routing collisions that cause existing routing methods to fail\cite{alfaresScalableCommodity2008, hopps2000analysis, li2024UCMP}.}

Many routing schemes were designed for the intra-DCs and rely on either feedback-driven reactivity\cite{jain2013B4Experience, ferguson2021OrionGoogles, song2023ConWeave} or randomized forwarding\cite{alfaresScalableCommodity2008, hopps2000analysis,li2024UCMP}. Both approaches suffer in inter-DC networks for two reasons. 
\ydy{First, \textit{\textbf{slow and outdated feedback signals}}: congestion signals traverse long paths, so reactive decisions may act on outdated information.
Second, \textit{\textbf{path heterogeneity and asymmetry}}: different from intra-DC links, topologically similar paths may differ greatly in propagation delay and link capacity in long-haul network while oblivious hashing or capacity-only metrics can misplace flows.
}

To illustrate, consider an inter-DC scenario (\figref{fig:moti-total}). From DC1 to DC8, there are six candidate routes (two high-, two medium-\nolinebreak[4], two low-capacity), and each capacity class contains one low-delay and one high-delay path. When RDMA traffic is sent between DC1 and DC8, we observe two effects. First, a capacity-centric policy (UCMP) concentrates traffic on the high-capacity/high-delay paths (e.g., the DC1–DC2 link shows 17\% utilization under UCMP vs. 6\% under ECMP), leaving lower-delay capacity underused. Second, ECMP’s random hashing can instead choose some low-delay links (e.g., DC1–DC6 and DC1–DC7 reach 30\% and 27\% utilization, respectively) while UCMP may avoid them entirely (0\% in \figref{subfig:moti-util}). These placement choices directly raise median and tail FCTs. These observations motivate a routing-centric design that fuses stable path quality with timely congestion signals to guide per-flow placement.

However, designing such a framework introduces three core challenges (details in \ref{subsec:challenges}):
\begin{enumerate}[label=,
    itemindent=-1.7em, 
    leftmargin=1.7em,topsep=0pt]
    \item \circled{C1} \textbf{Heterogeneous and asymmetric topology}: how can we define a compact ``path-quality'' score that captures both propagation delay and link capacity? (Solved in \ref{sec:path-quality})
    \item \circled{C2} \textbf{Slow and easily outdated congestion signals}: how can a datacenter interconnection (DCI) switch rapidly and robustly detect imminent congestion on inter-DC paths so that routing decisions remain effective despite long RTTs? (Solved in \ref{sec:cong-estimator})
    \item \circled{C3} \textbf{Simultaneous flow arrivals}: how can we avoid selection conflicts when many flows choose paths simultaneously? (Solved in \ref{subsec:selection-engine})
\end{enumerate}

To address these challenges, we present \name, a distributed \textbf{L}ong-haul \textbf{C}ost-aware \textbf{M}ulti-\textbf{P}ath routing framework for inter-DC RDMA.
\ydy{\name fuses a compact, control-plane precomputed path-quality score \(C_{\mathrm{path}}\) (encoding delay and capacity) with an integer-friendly on-switch congestion score \(C_{\mathrm{cong}}\) (instantaneous queue level, short-term trend, and persistence). The switch computes a fused cost per candidate, filters high-cost suffixes, and performs a diversity-preserving hash within the reduced set. }

\ydy{Importantly, \name is orthogonal to end-host congestion control and requires only modest upgrades to DCI switches. End hosts and intra-DC fabrics remain unchanged. We evaluate \name on a small-scale testbed and with large-scale NS-3 simulations against ECMP, UCMP reproduction, and several ablations. Across heterogeneous topologies and bursty workloads (including a 2,000 km inter-DC scenario), \name substantially reduces median and tail FCTs. We further present sensitivity and ablation studies to justify our parameter choices.}

\footnotetext{The propagation delay of 1000~km is 5~ms = $\frac{1000~\mathrm{km}}{2\times10^{8}~\mathrm{m/s}}$, where $2\times10^{8}~\mathrm{m/s}$ is the transmission speed of light in fiber\cite{li2025RevisitingRDMA}.}

\paragraph{Contributions} This paper makes three main contributions:
\begin{itemize}[topsep=0pt]
  \item We introduce \name, a distributed cost-fusion routing framework in long-haul inter-DC network that enables fast routing decisions with low deployment cost.
  \item We develop a compact path-quality representation and an on-switch congestion estimator that permit accurate comparison of heterogeneous inter-DC paths.
  \item We demonstrate that, in testbed and large-scale NS-3 experiments across realistic heterogeneous topologies (under the 2000 km inter-DC scenario), \name reduces median and tail FCT slowdown by up to 76\% and 64\%, respectively, compared to the SOTA routing baselines.
\end{itemize}

The rest of the paper is organized as follows. \ref{sec:background} presents background and challenges. \ref{sec:design} details the design. \ref{sec:feasibility} gives feasibility analysis. \ref{sec:implementation} describes deployment considerations. \ref{sec:evaluation} evaluates the system. \ref{sec:discussion} discusses limitations and future work. \ref{sec:relatedWork} reviews related work and \ref{sec:conclusion} concludes.

\section{Background \& Challenges}
\label{sec:background}

\subsection{Long-Haul RDMA Background}

Remote Direct Memory Access (RDMA) is widely used in clouds because it bypasses the kernel and offloads the network stack to RNICs, delivering very low latency, high throughput, and low CPU overhead \cite{zhu2015CongestionControl, guo2016RDMACommodity}. RDMA workloads are latency-sensitive. They favor in-order delivery and they suffer when packets are reordered.

Operators increasingly deploy RDMA across geographically distributed datacenters to support geo-replicated storage, distributed ML training, and remote memory services \cite{bai2023EmpoweringAzure, gangidi2024RDMAEthernet, gao2021WhenCloud}. Cross-region RDMA preserves RNIC-level performance benefits and simplifies application design. At the same time, it exposes RDMA flows to wide-area conditions that stress both transport and routing.

However, inter-DC links differ sharply from intra-DC links. Typical intra-DC propagation delays are on the order of microseconds. Inter-DC propagation delays range from milliseconds up to hundreds of milliseconds. Provisioned capacities across inter-DC links are heterogeneous (tens to hundreds of Gbps). Topologies are sparser and less regular than leaf–spine fabrics. These differences change how routing choices affect performance. Below we summarize the key distinctions and their routing implications.

\textbf{1) Large RTTs and outdated feedbacks.} Inter-DC links span hundreds to thousands of kilometers. One-way delays grow from microseconds to milliseconds and RTTs can be tens to hundreds of milliseconds. Long RTTs make controller- or host-driven feedback slow to reflect current congestion, so routing decisions that rely on recent global signals become outdated.

\textbf{2) Path asymmetry and heterogeneous topology.} Inter-DC topologies are sparser and less regular than intra-DC fabrics. Candidate routes that look equivalent at the topology level often show asymmetric delay–capacity trade-offs. Oblivious hashing (e.g., ECMP) or uniform-cost choices ignore these asymmetries and can systematically place flows on suboptimal paths.

These differences imply two requirements for inter-DC RDMA routing. First, routing must explicitly account for both propagation delay and provisioned capacity when ranking paths. Second, routing must use timely signals that indicate imminent congestion (so decisions remain useful despite long RTTs). We use these requirements to motivate the design of our cost-fusion, on-switch scoring, and diversity-preserving selection mechanisms (\ref{sec:design}).

\subsection{Existing Routing Approaches and Their Gaps}
\label{sec:routing-gaps}
Existing DC routing and traffic-engineering techniques are mature but have gaps when applied to long-haul RDMA traffic. Equal-Cost Multipath (ECMP\cite{alfaresScalableCommodity2008, hopps2000analysis}) is simple and widely deployed but hashes obliviously and ignores capacity/delay asymmetry. Weighted schemes (e.g., WCMP\cite{zhou2014WCMPWeighted}) incorporate static weights to address asymmetry, yet they are based on slow topology information and lack timely congestion awareness. Utility/capacity-aware approaches (e.g., UCMP\cite{li2024UCMP}) blend bandwidth and latency considerations but were designed for specific architectures (e.g., reconfigurable DCNs) and often rely on assumptions—like circuit wait costs, that do not hold in conventional WANs. Centralized SDN traffic engineering (e.g., B4-style controllers\cite{jain2013B4Experience, singh2015JupiterRising, yap2017TakingEdge, ferguson2021OrionGoogles, zhang2018BDSCentralized, zhang2021BDSInterDatacenter}) can optimize global utilization but incurs control-plane latency that makes it hard to react to fast congestion in high-RTT environments. Flowlet or packet-spraying techniques\cite{kandula2007DynamicLoad} improve utilization but risk RDMA reordering or require host/ASIC changes.

In short, most prior schemes either (a) ignore static path heterogeneity, (b) depend on slow feedback, or (c) require host or heavy switch changes. These gaps map directly to our design challenges \circled{C1}-\circled{C3} and motivate a routing approach that fuses slow control-plane path quality with timely, hardware-friendly on-switch congestion cues while preserving RDMA constraints (see \ref{sec:design}).

\subsection{Key Challenges and Solutions}
\label{subsec:challenges}
The background above can be summarized into three challenges that any practical inter-DC RDMA routing design must address. 

\circledTitle{C1}{How can we define a concise ``path quality'' representation that captures both propagation delay and link capacity? (Solved in \ref{sec:path-quality})}

Inter-DC topologies exhibit substantial heterogeneity: different candidate paths vary widely in propagation delay and in provisioned capacity. A routing metric must compress these partly static, partly slow-varying attributes into a form that switches can compare at line rate.

The path representation should (i) jointly reflect propagation delay and nominal capacity, (ii) be stable enough to be computed or normalized by the control plane and installed on the switch as compact per-path scores, and (iii) avoid expensive per-packet arithmetic on the data plane (i.e., the data plane should only do lookups and integer comparisons).

If path heterogeneity is ignored, capacity-aware methods may choose high-bandwidth but high-latency routes (hurting FCT), while latency-only choices underutilize available capacity. A concise, precomputed Path Quality score enables fast on-switch comparisons and informed trade-offs between delay and throughput.

\circledTitle{C2}{How can a DCI switch rapidly detect imminent congestion on inter-DC paths so that routing decisions remain effective despite long RTTs? (Solved in \ref{sec:cong-estimator})}

In inter-DC links, conventional congestion feedback (ECN) is delayed by large propagation times. Moreover, instantaneous queue length confuse transient bursts with sustained growth. As a consequence, signals are often too outdated for timely route decisions, while naive use of instantaneous samples causes noisy, oscillatory behavior.

A practical routing-oriented congestion signal must (i) be responsive to imminent queue buildup, (ii) suppress high-frequency noise to avoid undue re-routing, (iii) be representable as a compact quantized value (e.g., an 8-bit score), and (iv) be computable in the data plane using only hardware-friendly primitives.

Without such timely and implementable congestion sensing, switches either make decisions too late, which causes transient tail-latency spikes, or overreact to bursts and cause frequent path churn. Both outcomes degrade flow completion times and overall system predictability.

\circledTitle{C3}{How can we efficiently avoid selection conflicts when many flows make routing choices at the same time? (Solved in \ref{subsec:selection-engine})}

Inter-DC traffic often involves bursts of new flows that start near-simultaneously. If each new flow independently selects the currently cheapest path, many flows may concentrate on the same next-hop (a selection cascade, we call it herd effect), quickly saturating its egress queue and producing severe short-term tail latency.

A deployable mitigation must (i) rely on atomic, low-cost operations (register add/sub, comparisons), and (ii) preserve path diversity (e.g., by filtering high-cost candidates then randomizing among the low-cost set).

Without an efficient and bounded-state mechanism to prevent selection cascades, locally optimal per-flow choices will collectively create global congestion spikes and tail-latency degradation. Practical herd mitigation is therefore essential for robust routing in high-concurrency inter-DC environments.


\parab{Solutions.}
\name addresses the above three challenges with the following solutions for a practical inter-DC routing system. 

\begin{enumerate}[
    leftmargin=1.7em,
    topsep=0pt 
    ]
    \item Providing a compact, deployable path-quality representation. (See \ref{sec:path-quality})
    \item Designing a timely, data-plane friendly congestion estimator. (See \ref{sec:cong-estimator})
    \item Enabling herd mitigation with diversity-preserving selection under simultaneous flow arrivals. (See \ref{subsec:selection-engine})
\end{enumerate}

\section{\name Design}
\label{sec:design}

\subsection{Design Overview}
\label{subsec:design-overview}

\begin{figure}[t]
  \centering
  \includegraphics[width=\linewidth]{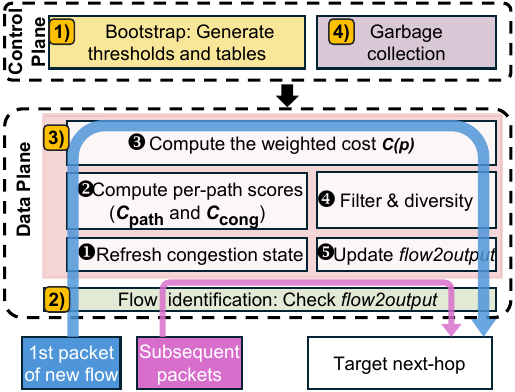}
  \caption{\name architecture overview.}
  \Description{\name architecture overview.}
  \label{fig:overview}
\end{figure}

\subsubsection{High-Level Abstraction}
\label{subsubsec:highLevelIdea}
\name makes per-flow next-hop decisions by fusing a control-plane view of path quality with on-switch congestion signals. Concretely, for a candidate path $p$ we compute an integer cost,
\begin{align}
\label{eq:weightedCost}
C(p) \;=\; \alpha \cdot C_{\mathrm{path}}(p) \;+\; \beta \cdot C_{\mathrm{cong}}(p),
\end{align}
where $C_{\mathrm{path}}$ is a precomputed control-plane score that encodes propagation delay and provisioned capacity, and $C_{\mathrm{cong}}$ is a congestion score derived from instantaneous queue level, short-term trend, and a persistence penalty. The switch picks the final egress from a low-cost candidate set.
The abstraction directly targets the three challenges identified in \ref{subsec:challenges}:

\textbf{Addressing \circled{C1} heterogeneous, asymmetric topologies.}
We separate slowly-varying path attributes from transient congestion by precomputing a compact per-path quality score in the control plane (\ref{sec:path-quality}). Encoding delay and provisioned capacity into a score allows the data plane rapidly compare heterogeneous paths without global queries.

\textbf{Addressing \circled{C2} slow and easily outdated congestion signals.}
Rather than relying on end-to-end or controller-roundtrip feedback, each DCI switch maintains on-switch signals: a quantized instantaneous queue level, short-term trend accumulator, and a duration counter (\ref{sec:cong-estimator}). These signals focus the decision on local queue growth and are normalized to be robust to sampling noise and long RTTs.

\textbf{Addressing \circled{C3} many simultaneous flows and herd effects.}
To avoid simultaneous choices collapsing onto the same low-cost path, \name applies a two-stage selection: (i) filter out the high-cost candidate paths, and (ii) perform hash-based selection within the reduced, low-cost set (\ref{subsec:selection-engine}).

\subsubsection{Runtime Workflow}
\figref{fig:overview} provides a high-level overview of \name.

\paragraph{1) DCI Switch Bootstrap.}
At switch initialization time \name installs a small set of tables and threshold vectors that the data plane uses for fast mapping and normalization:

\textbf{Link capacity thresholds.} A small vector of increasing link capacity thresholds (e.g., $N=10$ classes) is created: each class boundary is proportional to a configured link capacity. These thresholds map link rates into a discrete link score lookup.

\textbf{Queue thresholds.} The switch divides its per-port egress buffer capacity into levels and records per-level thresholds. These thresholds are used to map instantaneous queue bytes to a quantized queue level $Q$.

\textbf{Level score table.} A linear mapping from level index to a 0--255 score is precomputed. This avoids per-packet floating computation.

\textbf{Trend normalization tables.} For each coarse link-rate bucket (e.g., 25/100/400 Gbps), a small per-level trend threshold vector is created. These tables normalize the raw trend accumulator into a trend level $T$. If a rate bucket is not present at initialization the data plane can create a small normalized table on-demand from the link rate.

\begin{figure}[t]
  \centering
  \includegraphics[width=1\linewidth]{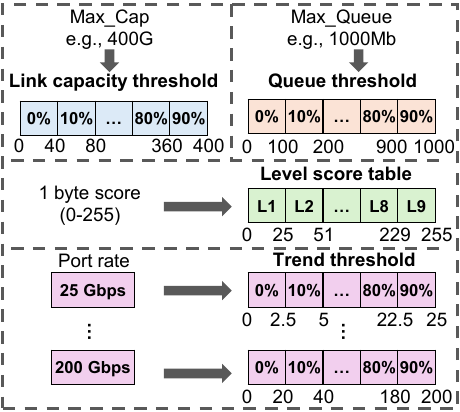}
  \caption{Switch bootstrap tables and mappings. Control plane installs a small set of vectors.}
  \Description{Switch bootstrap tables and mappings. Control plane installs a small set of vectors.}
  \label{fig:bootstrap}
\end{figure}

These compact data structures (a few small vectors and lookup tables per switch) are sized to fit on programmable switch memory and to be installed/updated by the control plane as link or provisioning information changes (see \figref{fig:bootstrap}).

\paragraph{2) Flow Identification.} On packet arrival the switch forms a flow identifier (e.g., a five-tuple hash). If the packet belongs to an \emph{established} flow (a $flow2output$ mapping exists), the switch refreshes the flow's last-seen timestamp and forwards the packet via the previously chosen egress. This guarantees path consistency and prevents out-of-order packets\cite{huang2025FastScalable}.

\paragraph{3) Flow Routing.} If the packet is the first packet of a flow, the switch executes the full \name decision path:

\textbf{Refresh congestion state} (\ding{202}). It invokes a light-weight monitor to sample per-port queue depth and update the short-term trend estimator. This step updates three signals for each candidate port:
(1) $Q$: queue occupancy mapped to a level via preinstalled thresholds;
(2) $T$: short-term trend obtained via a shift-based EWMA, $T = T_{\text{old}} - (T_{\text{old}}\gg K) + (\Delta \gg K)$, 
  where $\Delta$ is the queue-byte delta between samples, $K$ is an integer (e.g., 3), and $\gg$ denotes a right-bit-shift normalization;
(3) $D$: a duration (persistence) penalty that accumulates when $Q$ stays above a high-water mark and decays otherwise.

\textbf{Compute per-path scores} (\ding{203}). For each candidate path computes:
\begin{itemize}
  \item $delayScore$ via a shift-based mapping;
  \item $linkCapScore$ via a control-plane installed capacity-class lookup (data plane compares configured link capacity against threshold table and returns a score);
  \item $C_{\mathrm{path}}$ by combining $delayScore$ and $linkCapScore$ with integer weights and a right-shift normalization;
  \item $C_{\mathrm{cong}}$ by combining quantized $Q,T,D$ with integer weights and a right-shift normalization.
\end{itemize}

\textbf{Compute the weighted cost $C(p)$} (\ding{204}). Compute the weighted cost of each path with \eqnref{eq:weightedCost}.

\textbf{Filter and diversity-preserving selection} (\ding{205}). Sort candidate paths by the fused cost $C(p)$. Remove the high-cost suffix (paths above a cut), keep a reduced candidate set (we use the top 50\% by cost in our implementation), and perform a hash-based ECMP selection within that reduced set to pick the final egress. 

\textbf{Update flow2output mapping} (\ding{206}).
The selected mapping is then recorded in a flow table so that subsequent packets of the flow follow the same egress.

\paragraph{4) Garbage Collection.}
Per-flow consistency is necessary to avoid reordering and ensure stable path utilization. \name therefore maintains a bounded flow cache that maps a flow identifier to the chosen egress and a last-seen timestamp.

\textbf{Flow cache entry and operations.}
Each entry contains (1)Flow ID, (2) \texttt{outDevIdx}: chosen egress port/index, and (3) \texttt{lastSeen}: last packet arrival time.
On packet arrival an established flow entry is refreshed and the packet forwarded via the recorded egress. Only the first packet of a flow executes the full cost computation and selection.

\textbf{Garbage collection.}
A periodic garbage collection evicts entries whose \texttt{lastSeen} exceeds a configured idle timeout (e.g., a fraction of RTT-based shortTimeout or a conservative fixed value). This keeps the flow cache bounded and prevents outdated mappings from persisting indefinitely.

Importantly, the storage overhead of \name is \emph{small}, a 50k-entry simultaneous flow cache requires only 1.2 MB (see \ref{sec:feasibility} for details).

\subsection{Compact Control-Plane Path-Quality Representation}
\label{sec:path-quality}

Inter-DC topologies exhibit largely static but heterogeneous attributes (propagation delay and provisioned capacity) that should be respected by any path-selection policy. \name separates these slowly-varying, control-plane-friendly attributes from fast on-switch signals by precomputing a compact per-path \emph{path-quality} score $C_{\mathrm{path}}\in[0,255]$ and installing it as a small table on each DCI switch.

The control plane obtains per-link one-way propagation delay and configured link capacity, maps each metric to a score, and fuses them with integer weights:
\begin{align}
pathScore &= w_{dl}\cdot \mathrm{delayScore}(p) + w_{lc}\cdot \mathrm{linkCapScore}(p), \nonumber \\
C_{\mathrm{path}}(p) &= \min\Bigl(pathScore \gg S_{\mathrm{path}},\,255\Bigr). 
\end{align}
The mapping functions are deliberately simple and integer-only. As shown in \algref{alg:delay-to-score} and \algref{alg:linkCap-to-score}, delayScore linearly maps one-way delay to 0--255 (saturating at a configured maximum, e.g., 32, 64 ms), and linkCapScore maps link rate into a small number of classes via preinstalled thresholds.

\begin{algorithm}[t]
\SetAlgoLined
\DontPrintSemicolon
\KwIn{one\_way\_delay}
\KwOut{delayScore in $[0,255]$}
\BlankLine
MAX\_DELAY = 32 \tcp*{configured saturation point (ms)}
SHIFT = 5 \tcp*{right-shift equivalent to dividing by MAX\_DELAY\_MS}
\If{one\_way\_delay >= MAX\_DELAY}{
  \Return 255 \tcp*{at worst score}
}
delayScore $\leftarrow$ (one\_way\_delay * 255) $\gg$ SHIFT)\;
\Return delayScore
\caption{CalcDelayCost: saturating, shift-based mapping from delay to delayScore.}
\label{alg:delay-to-score}
\end{algorithm}

\begin{algorithm}[t]
\SetAlgoLined
\DontPrintSemicolon
\KwIn{linkCap, linkCapThresholds[0..N-1], levelScore[0..N-1]}
\KwOut{linkCapScore in $[0,255]$}
\BlankLine
\For{$i \leftarrow N-1$ \KwTo $0$ \textbf{By} $-1$}{
  \If{linkCap $\ge$ linkCapThresholds[i]}{
    \Return 255 - levelScore[i] \tcp*{higher capacity $\Rightarrow$ smaller cost}
  }
}
\Return 255
\caption{CalcLinkCapCost: link capacity-class lookup mapping link capacity to linkCapscore.}
\label{alg:linkCap-to-score}
\end{algorithm}

\subsection{Realtime, On-Switch Congestion Estimator}
\label{sec:cong-estimator}

Timely and noise-robust congestion signals are central to \name's effectiveness in long-RTT environments. \name generates a on-switch congestion score $C_{\mathrm{cong}}$ by fusing three signals: instantaneous queue level $Q$, a short-term trend level $T$, and a duration (persistence) penalty $D$.

\textbf{Instantaneous queue level $Q$.} The monitor samples per-port queue bytes and maps the sampled byte count into a discrete level via the preinstalled \texttt{qThresh} vector. The level index is then converted to a score via \texttt{levelScore}.

\textbf{Short-term trend $T$.} \name uses a shift-based EWMA-style accumulator:
\begin{align}
\label{eq:trendCalculate}
T = T_{\text{old}} - (T_{\text{old}} \gg K) + (\Delta \gg K).
\end{align}
The raw $\mathrm{trend}$ is mapped to a discrete trend level by comparing it to a normalization vector and converting the matched level to a score. Non-positive trends map to zero to focus reactions on growing queues.

\textbf{Duration penalty $D$.} A counter increases while $Q$ exceeds a high-water mark and decays when $Q$ is low. This persistence counter is right-shifted to produce a penalty score.

\textbf{Fusion into $C_{\mathrm{cong}}$.}
The three signals are combined with integer weights and a right-shift normalization:
\begin{align}
congScore &= w_{ql}\cdot Q + w_{tl}\cdot T + w_{dp}\cdot D, \\
C_{\mathrm{cong}}(p) &= \min\Bigl(congScore \gg S_{\mathrm{cong}},\,255\Bigr).
\end{align}

\textbf{Sampling and robustness.}
A lightweight monitor routine iterates over device ports at a modest cadence. Trend normalization uses the observed sampling interval when comparing the trend accumulator to per-rate thresholds, making $T$ robust to modest variations in sampling frequency. This design balances responsiveness to imminent queue growth with suppression of high-frequency noise.

\subsection{Diversity-Preserving Selection for Herd Mitigation}
\label{subsec:selection-engine}

To prevent simultaneous new-flow from choosing the same single low-cost port (called ``herd effect''), \name performs a two-stage selection: cost-based filtering followed by randomized selection within the reduced set.

\textbf{Two-stage selection.}
For a new flow the data plane computes the fused cost $C(p)$ for each candidate path $p$. The switch forms a vector of $(C(p),p)$ pairs, sorts them by cost (small $N$ so sorting is cheap), and removes the high-cost suffix. By default \name retains the lower half of candidates. From the remaining paths, the switch performs ECMP inside the low-cost subset.

\textbf{Fallbacks and corner cases.}
If all candidate paths are highly congested, \name falls back to selecting the minimum-cost path to avoid pointless randomization among uniformly bad choices. The per-flow mapping is then recorded in the local flow cache to preserve path consistency for subsequent packets.

\reviseadd{To B,C,D: CR-3}{
\textbf{Fault tolerance via data-plane fast-failover.}
\name handles link or port failures entirely in the data plane to avoid control-plane latency. The switch tracks port liveness status in real-time. If a packet matches a flow cache entry pointing to a failed port, the switch logic invalidates the entry on-the-fly and treats the packet as the ``first packet'' of a new flow. This triggers the path selection logic to immediately re-hash the flow to a remaining healthy candidate. We employ a lazy-update design: instead of the control plane performing costly batch updates to modify thousands of flow entries upon a failure, invalid entries are overwritten individually only when packets for those flows arrive. This ensures µs-scale recovery with zero instantaneous control-plane overhead.
}





\section{Analysis of Resource Cost}
\label{sec:feasibility}

Before describing a concrete implementation, we quantify \name's resource and decision compute requirements to demonstrate that the design is practical on modern DCI switches. We provide a parameterized accounting of per-port and per-flow storage, a conservative example deployment (48 ports, 50k-entry flow cache), and a breakdown of the \emph{per-new-flow} integer operations and table lookups required for the full cost computation and selection.

Importantly, \name performs the relatively expensive cost computation only once per new flow: subsequent packets of the same flow hit the local flow cache, incur a simple lookup, refresh the last-seen timestamp, and are forwarded via the recorded egress. Our accounting therefore focuses on the per-new-flow decision cost, roughly a few dozen table lookups and $O(m\log m)$ comparisons for $m$ candidate next-hops. The numbers below show that \name's working set and its per-new-flow compute comfortably fit within typical programmable-switch budgets.

\textbf{Per-element sizes.}
We assume the following conservative storage sizes typical in switch registers:
(1) 32-bit integer fields (e.g., \texttt{queueCur}, \texttt{queuePrev}, \texttt{trend}, \texttt{durCnt}): 4 bytes (B) each.
(2) 64-bit timestamps (e.g., \texttt{lastSample}, \texttt{lastSeen}): 8 B each.
(3) Per-path or per-level 8-bit scores: 1 B each (stored in table entries).

\textbf{Per-port and per-flow memory overhead.}
\begin{align}
\text{Per-port bytes} &= \underbrace{4}_{\text{queueCur}} + \underbrace{4}_{\text{queuePrev}} + \underbrace{4}_{\text{trend}} + \underbrace{4}_{\text{durCnt}} + \underbrace{8}_{\text{lastSample}} \nonumber \\
&= 24\ \text{B/port}, \nonumber \\
\text{Per-flow bytes} &= \underbrace{8}_{\text{flowId}} + \underbrace{4}_{\text{portIdx}} + \underbrace{8}_{\text{lastSeen}} = 20\ \text{B/flow}. \nonumber
\end{align}

\textbf{Demonstration.}
Consider a DCI switch with 48 ports and a bounded flow cache sized for 50,000 entries. Using the formulas above:

\begin{itemize}
  \item \textbf{All port cache:} $24~\text{B/port} \times 48~\text{ports} = 1152~\text{B}$.
  \item \textbf{All flow cache:} $24~\text{B/flow} \times 50{,}000~\text{flows} =1.2~\text{MB}$. 
  \item \textbf{Control tables:} bandwidth thresholds and levelScore for $N=10$ classes: approximately a few dozen bytes each. Per-path $C_{\mathrm{path}}$ table size depends on the number of installed paths $P$, e.g., $P=10$K paths $\approx$ 10 KB for scores.
\end{itemize}

These totals (roughly 1.2\,MB) are well within typical on-switch memory budgets (and can be kept smaller if resources are constrained with some methods\cite{chen2024PreciseData}).

\textbf{Per-new-flow computational cost.}
Let $m$ be the number of candidate next-hops (typical $m\in[2,8]$). For each candidate the pipeline performs:
\begin{itemize}
  \item 2–4 table lookups (bandwidth class, levelScore, trend thresholds),
  \item a handful of integer ops (compute delayScore, combine weights: 8–12 adds/shifts),
  \item compare operations to form sort keys.
\end{itemize}
A conservative per-candidate estimate is 15 integer primitives, thus for $m=6$ the cost is 90 primitives plus a small sorting cost (for $m=6$, sorting requires on the order of $m\log_2 m = 6\times2.6 \approx 15$ comparisons). Total primitive count 105 integer operations for a new-flow decision, which is trivial for modern ASIC pipelines or programmable switch.
\section{Implementation}
\label{sec:implementation}

We implemented a prototype of \name on Tofino programmable switches. \reviseadd{To A, C: CR-2}{This design requires no new ASIC features or complex operations, ensuring it fits within the resource constraints of modern hardware.}
Only DCI (inter-DC) edge switches require an upgrade. End hosts and the intra-DC fabric remain \emph{unchanged}, enabling low-risk, incremental deployment.

\textbf{Dataplane requirements.} \name targets commonly available dataplane primitives: a few compact lookup tables (per-path $C_{\mathrm{path}}$, bandwidth thresholds and level-score vectors), a small set of 32-bit per-port registers (queue, trend, duration) and a bounded per-flow cache, and integer-only operations (adds, right-shifts, comparisons) plus cheap sorts over a small candidate set.

\textbf{Control-plane provisioning.} The controller performs only slow-path work: installing per-path $C_{\mathrm{path}}$ scores and threshold vectors, pushing conservative default weights (e.g., $(\alpha,\beta)=(3,1)$) for operator tuning, and collecting lightweight telemetry (per-port queue levels, flow-cache occupancy) for verification. 

\textbf{Incremental rollout and safe fallbacks.} LCMP supports partial upgrades: upgraded DCIs apply LCMP locally while legacy devices continue normal forwarding. Decisions are local next-hop choices and do not require new packet headers or remote upgrades. If LCMP tables are missing or outdated, or all candidates are uniformly poor, switches fall back to ECMP.

\textbf{Compatibility with transport.} LCMP is orthogonal to end-host CC: it requires \emph{no} RNIC or host-stack changes and interoperates with DCQCN, HPCC, etc.


\section{Evaluation}
\label{sec:evaluation}

\begin{figure}[t]
  \centering
  \begin{subfigure}[t]{1.4954 in}
    \centering
    \includegraphics[width=\linewidth]{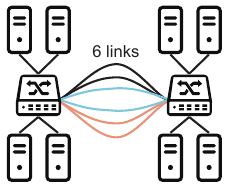}
    \caption{Testbed: 8-DC topology.}
    \label{fig:topo-testbed}
  \end{subfigure}\hfill
  \begin{subfigure}[t]{1.6035 in}
    \centering
    \includegraphics[width=\linewidth]{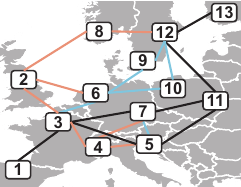}
    \caption{BSONetwork: real-world Europe-spanning topology.}
    \Description{BSONetwork: real-world Europe-spanning topology.}
    \label{fig:topo-bso}
  \end{subfigure}
  \caption{Topologies used in evaluation.}
  \label{fig:topologies}
\end{figure}

\ydy{Our evaluation across a small-scale emulated testbed and large-scale NS-3 simulations reveals the following key findings:}
\begin{enumerate}[leftmargin=*,itemsep=2pt,topsep=3pt]
  \item On the 8-DC testbed (\figref{fig:topo-testbed}) \name reduces median and tail FCT slowdown by up to 76\% and 64\%, respectively, compared to the SOTA method UCMP (\ref{subsec:smallTestbed}).
  \item For endpoint pairs with many candidate routes under the \emph{2000 km inter-DC scenario}, \name delivers clear benefits: median FCT improves by $7\%{-}11\%$ and P99 by $15\%{-}18\%$ versus ECMP (even larger improvements versus UCMP) (\ref{subsec:largeScaleNS3}).
  \item Improvements persist across realistic workloads and across several RDMA-capable CCs: \name reduces median FCT slowdown by $32\%{-}35\%$ and $74\%{-}75\%$, and P99 slowdown by $39\%{-}45\%$ and $40\%$ compared to ECMP and UCMP, respectively (\ref{sec:deep-dive}).
\end{enumerate}

\subsection{Small-Scale Emulated Testbed Experiments}
\label{subsec:smallTestbed}

\paragraph{Testbed topology.} As shown in \figref{subfig:moti-porCor}, we use a 8-DC topology and each DC is a small leaf–spine fabric (1 DCI switch, 2 spine switches, 4 leaf switches, and 16 servers). Servers attach to leaf switches via a single NIC. All intra-DC links run at 100 Gbps and use a 1 \textmu s propagation delay. To avoid artificial bottlenecks inside a DC, links between DCI switches and spine switches are set to 400 Gbps. The inter-DC link capacities are set to 40 Gbps, 100 Gbps, and 200 Gbps and propagation delays are set from 5 ms to 250 ms.

\paragraph{Workloads}
Here we use a realistic DCN workload Web Search\cite{zhu2015CongestionControl}. We synthesize an all-to-all inter-DC traffic pattern by randomly pairing senders and receivers between DC1 and DC8. 

\paragraph{Baselines} 

We compare \name against three practical baselines representing widely deployed, capacity-aware, and SOTA WAN traffic engineering (TE) strategies. ECMP\cite{alfaresScalableCommodity2008, hopps2000analysis} is the common default routing scheme in DCNs, which hashes flows across paths deemed to have equal cost. UCMP\cite{li2024UCMP} is a recent scheme proposed for reconfigurable datacenter networks that combines circuit-waiting latency and link capacity considerations into a unified cost to guide path selection. \reviseadd{To A,E: CR-1.2}{
RedTE\cite{gui2024RedTEMitigating} represents the SOTA in distributed WAN TE. It leverages multi-agent reinforcement learning to dynamically adjust traffic splitting ratios at edge routers to mitigate sub-second traffic bursts.}

\paragraph{Metrics} Our primary metric is \textbf{FCT slowdown}\cite{li2019HPCCHigh}. It means a flow’s actual FCT normalized by its ideal FCT. Ideal FCT is the FCT of the same flow when run alone in the network with the shortest propagation delay in its topology, which isolates queueing effects due to multiplexing. We repeat the experiment three times.

\paragraph{Setup.}
We build a small-scale emulation consisting of 9 servers (see \figref{fig:topo-testbed}), which is simplified form \figref{subfig:moti-porCor}. 4 machines are grouped behind a DCI switch and act as DC1, and another 4 serve as DC8. The remaining host runs Mininet\cite{zhang2024FedRDMACommunicationefficient} solely to emulate the long-haul propagation delays and link capacities between the DCs. This setup validates protocol correctness and logic flow.
\reviseadd{To A, C: CR-2}{Since high-speed RNICs were unavailable for this specific testbed, we utilized SoftRoCE on standard Ethernet NICs to emulate the RoCEv2 transport stack and used perftest for traffic generation.}
DCQCN\cite{zhu2015CongestionControl} is used as the default CC. We run the workload at 30\%, 50\% and 80\% load(i.e., light, medium and heavy load).

\begin{figure}[t]
  \centering
  \includegraphics[width=\linewidth]{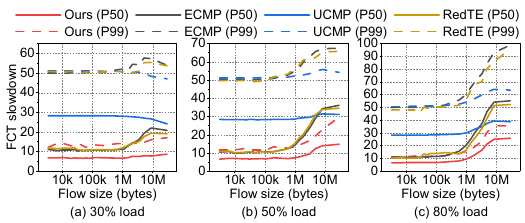}
  \caption{Median and tail FCT slowdown for \textit{Web Search} on the testbed topology under 30\%, 50\%, 80\% load.}
  \Description{Median and tail FCT slowdown for \textit{Web Search} on the testbed topology under 30\%, 50\%, 80\% load.}
  \label{fig:exp-testbed}
\end{figure}

\paragraph{Results.}  
As shown in \figref{fig:exp-testbed}, across three loads \name reduces median FCT slowdown by $36\%-41\%$, $76\%$ and $36\%-54\%$ compared to ECMP, UCMP, and RedTE, respectively. 
For P99 tail latency, \name achieves reduction of  $56\%-68\%$, $45\%-64\%$, and $73\%-77\%$ against these baselines. 
These improvements arise because \name avoids ECMP's random placement on high-delay links and UCMP's capacity-only bias by fusing path quality with on-switch congestion signals. \reviseadd{To A, E: CR-1.2}{Notably, RedTE exhibits performance similar to ECMP in this scenario. Its 100ms control loop is too coarse to capture the µs-scale micro-bursts of RDMA traffic, causing it to effectively degenerate to static hashing.}

\begin{figure}[htbp]
  \centering
  \includegraphics[width=\linewidth]{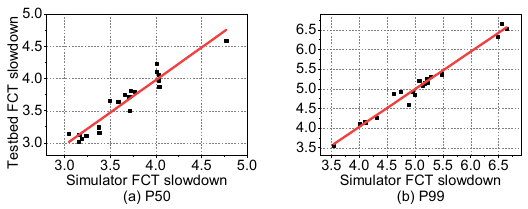}
  \caption{[Simulator fidelity] NS-3 vs testbed FCT slowdown.}
  \Description{Simulator fidelity: NS-3 vs testbed FCT slowdown.}
  \label{fig:sim-fidelity}
\end{figure}

\paragraph{Simulator fidelity.}
\figref{fig:sim-fidelity} compares FCT slowdown measured on our testbed and in the NS-3 simulator under 30\% load with the same setting. The line shows the near-linear correlation between them (the Pearson correlation values are 95\% for P50 and 97\% for P99), which validates NS-3 as a faithful platform for the larger-scale experiment. Consequently, all remaining experiments use NS-3 results.

\subsection{Large-Scale NS-3 Simulations}
\label{subsec:largeScaleNS3}

\paragraph{Real-world topology}
\figref{fig:topo-bso} provides a realistic European network topology (\textit{BSONetworkSolutions}) drawn from the Internet Topology Zoo\cite{knight2011TopologyZoo}. This topology contains backbone, customer and transit links across regions and therefore captures realistic heterogeneity in both delay and capacity.
There are \emph{13 DCs} and we set inter-DC propagation delays to 1 ms (for 200 km), 5 ms (for 1000 km) and 10 ms (for 2000 km), and increase switch buffer sizes to 6 GB for the long distances \cite{li2025RevisitingRDMA} to reflect long-haul provisioning and to satisfy PFC headroom requirements for RDMA traffic.

\paragraph{Workloads}
In addition to WebSearch\cite{roy2015SocialNetwork}, we use two more realistic DCN workloads in our experiments, which is Facebook Hadoop\cite{alizadeh2010DCTCP}, and Alibaba Storage\cite{li2019HPCCHigh}. 
For each workload we synthesize an all-to-all inter-DC traffic pattern by randomly pairing senders and receivers \emph{across all DCs}. We also vary the offered load to achieve average link utilizations of 30\%, 50\% and 80\%.

\paragraph{Baselines} 
The same methods used in testbed: ECMP, UCMP \reviseadd{To A,E: CR-1.2}{and RedTE}.

\subsubsection{System-Wide Validation: Aggregate FCT for All-to-All Inter-DC Flows}
\paragraph{Setup} 
We use NS-3 for simulations under 30\%, 50\%, and 80\% traffic loads. 
\reviseadd{To B: R-2}{
We utilize the WebSearch here as the representative benchmark. As detailed later in \ref{sec:deep-workload}, \name maintains consistent performance trends across other diverse workloads. 
}
All 13 DCs participate in an all-to-all inter-DC traffic matrix. 

\begin{figure}[htbp]
  \centering
  \includegraphics[width=\linewidth]{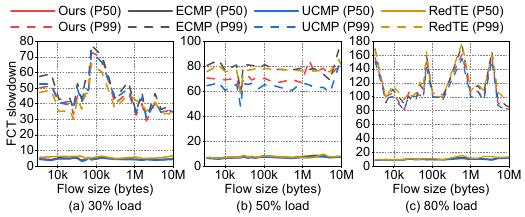}
  \caption{[System-wide validation] Median and tail FCT slowdown across all inter-DC flows at 30\%, 50\% and 80\% loads.}
  \Description{Median and tail FCT slowdown across all inter-DC flows at 30\%, 50\% and 80\% loads.}
  \label{fig:exp-agg}
\end{figure}

\paragraph{Results} 
As shown in \figref{fig:exp-agg}, \name does not harm overall median performance and yields modest tail improvements. Compared to ECMP the median FCT slowdown is essentially unchanged across the three loads, while the P99 FCT falls by roughly $2\%-9\%$. 
Against UCMP, \name shows comparable tail reductions, though UCMP sometimes produces slightly lower medians by biasing towards high-capacity paths. 
\reviseadd{To A,E: CR-1.2}{Compared to RedTE, \name reduces P99 FCT by up to $54\%$}

\reviseadd{To B: R-3}{We observe that the system-wide gains in the realistic 13-DC simulation are more moderate compared to \figref{fig:exp-testbed}. This stems from two differences. First, path diversity: the 13-DC is sparser, where only $25.6\% (20/78)$ of node pairs have multiple candidate paths (vs. $57.1\% (16/28)$ in the testbed). Consequently, the significant gains on multi-path flows are diluted by the majority of single-path flows. Second, latency heterogeneity: the testbed configured extreme delay gaps (50×: 5ms vs. 250ms) to stress-test path selection, whereas the realistic topology has smaller delay gaps (10×: 1ms vs. 10ms).}

\subsubsection{Representative DC-Pair Case Study: (DC1, DC13)}

\paragraph{Setup.} To highlight \name's mechanism, We filter the same runs used above to extract flows between DC1 and DC13, which exhibit multiple candidate routes.

\begin{figure}[htbp]
  \centering
  \includegraphics[width=\linewidth]{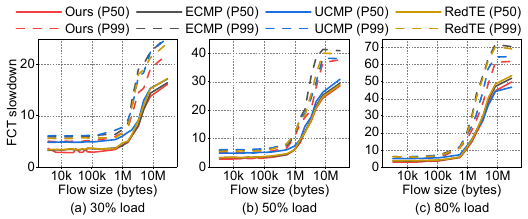}
  \caption{[DC-pair case study] Median and tail FCT slowdown for flows between DC pair (DC1, DC13) at 30\%, 50\% and 80\% loads.}
  \Description{Median and tail FCT slowdown for flows between DC pair (DC1, DC13) at 30\%, 50\% and 80\% loads.}
  \label{fig:exp-dcpair}
\end{figure}

\paragraph{Results.}
When we focus on a representative DC-pair with multiple candidate routes (DC1–DC13), \name’s benefits become clear in \figref{fig:exp-dcpair}. For flows between DC1 and DC13, \name reduces median slowdown by $7\%{-}11\%$ and P99 slowdown by $15\%{-}18\%$ relative to ECMP and RedTE. Versus UCMP the improvements are larger for medians (median slowdown drops by $25\%{-}30\%$) while tails fall by $13\%{-}16\%$. These focused improvements arise because DC1–DC13 runs have multiple viable next-hops with differing delay and capacity trade-offs: \name’s fusion of path-quality with on-switch congestion signals both (i) avoids systematically placing latency-sensitive flows on high-delay or high-capacity paths and (ii) mitigates transient herding on a single low-cost port, producing substantially better median and tail FCTs in multi-path inter-DC scenarios.

\subsection{Deep Dive}
\label{sec:deep-dive}
Having established system-wide behavior in the previous section, here we omit repeat aggregate results and focus on the representative DC pair (DC1, DC8) in the \figref{subfig:moti-porCor}. We will further demonstrate \name's robustness across realistic workloads and common CC algorithms.

\subsubsection{Workload Sensitivity}
\label{sec:deep-workload}
\paragraph{Setup.}
We run three DC workloads (Web Search, Facebook Hadoop, Alibaba Storage) at 30\% load using DCQCN as the default CC. 

\begin{figure}[htbp]
  \centering
  \includegraphics[width=\linewidth]{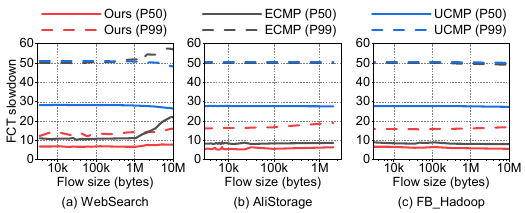}
  \caption{Workload sensitivity: median and tail FCT slowdown different three workloads.}
  \Description{Workload sensitivity: median and tail FCT slowdown different three workloads.}
  \label{fig:exp-workloads}
\end{figure}

\paragraph{Results.}
\figref{fig:exp-workloads} shows that, for \textit{Web Search} \name reduces median slowdown by $36\%$ and P99 slowdown by $58\%$ versus ECMP, and by $76\%$ (median) and $82\%$ (tail) versus UCMP. For \textit{Alibaba Storage} \name cuts median/tail by $32\%$/$68\%$ versus ECMP and by $80\%$/$68\%$ versus UCMP. For \textit{Facebook Hadoop} \name reduces median/tail by $26\%$/$69\%$ versus ECMP and by $78\%$/$69\%$ versus UCMP. These results show that median improvements primarily stem from \name respecting path-quality (avoiding high-delay, high-capacity routes), while the large tail reductions come from the on-switch congestion estimator and diversity-preserving selection.

\paragraph{Takeaway.}
\name’s benefits are robust to realistic variations in flow-size distributions: improvements in both p50 and P99 persist across workloads. 

\subsubsection{Congestion-Control Orthogonality}
\paragraph{Setup.}
we evaluate \name’s interaction with multiple end-host CCs: DCQCN (shown in \figref{fig:exp-testbed}), HPCC, TIMELY and DCTCP. All experiments use the Web Search workload at 30\% load.

\paragraph{Results.}

\begin{figure}[htbp]
  \centering
  \includegraphics[width=\linewidth]{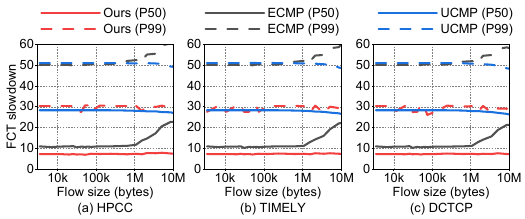}
  \caption{Congestion-control orthogonality: median and tail FCT slowdown under different CCs.}
  \Description{Congestion-control orthogonality: median and tail FCT slowdown under different CCs.}
  \label{fig:deep-cc}
\end{figure}

Across all tested CC algorithms, \figref{fig:deep-cc} shows that, \name delivers highly consistent benefits: \name reduces median FCT slowdown by $32\%{-}35\%$ and $74\%{-}75\%$, and P99 slowdown by $39\%{-}45\%$ and $40\%$ compared to ECMP and UCMP, respectively. The numbers are stable across the four CCs we tested (DCQCN earlier, plus HPCC, TIMELY and DCTCP here), indicating that \name's improvements are largely orthogonal to the choice of end-host CC.

This pattern has two implications. First, it shows \name is plug-and-play: operators can deploy \name alongside existing CCs and expect similar improvements without changing host stacks. Second, the similarity across CCs suggests a broader lesson: many CC algorithms developed for intra-DCs rely on timely feedback and small RTTs, assumptions that weaken in an inter-DCs (large-RTT) . Consequently, future CC research for Inter-DCs should (i) revisit feedback mechanisms to provide faster, more informative signals over long RTTs, and (ii) explore cross-layer designs that let routing and CC share concise path-quality and imminent-congestion costs. These directions would complement routing-centric solutions like \name and further improve both FCT performance in multi-DCs.

\paragraph{Takeaway.}
These results confirm \name’s orthogonality: operators can adopt \name without changing RNICs or transport protocols and still obtain consistent median/tail reductions. This makes \name a low-risk, deployable addition to current inter-DC stacks.

\section{Sensitivity Analysis and Discussion}
\label{sec:discussion}

\begin{figure*}[htbp]
  \centering
    \begin{subfigure}[T]{0.23\textwidth}  
        \includegraphics[width=\linewidth]{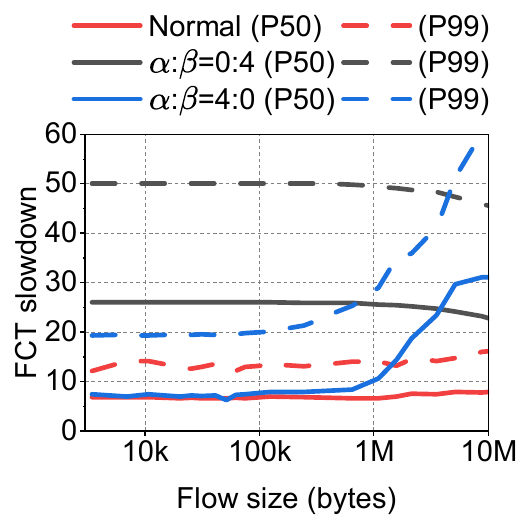} 
        \caption{\centering [Ablation analysis]}
        \label{fig:discuss-ablation}
    \end{subfigure}
    \begin{subfigure}[T]{0.23\textwidth}  
        \includegraphics[width=\linewidth]{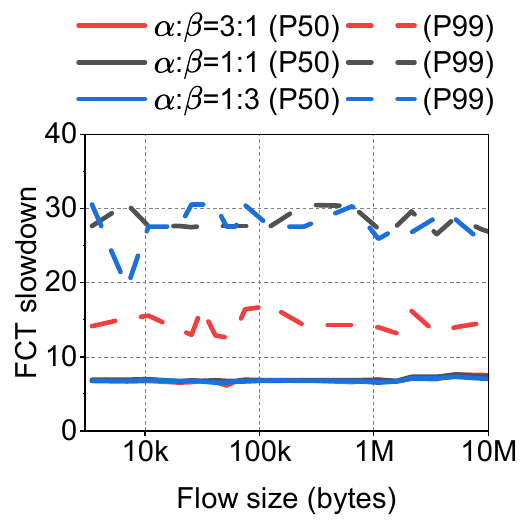} 
        \caption{\centering [Global weight analysis] Weight tuples $(\alpha,\beta)=(3,1),(1,1),(1,3)$}
        \label{fig:discuss-global-weights}
    \end{subfigure}
    \begin{subfigure}[T]{0.23\textwidth}  
        \includegraphics[width=\linewidth]{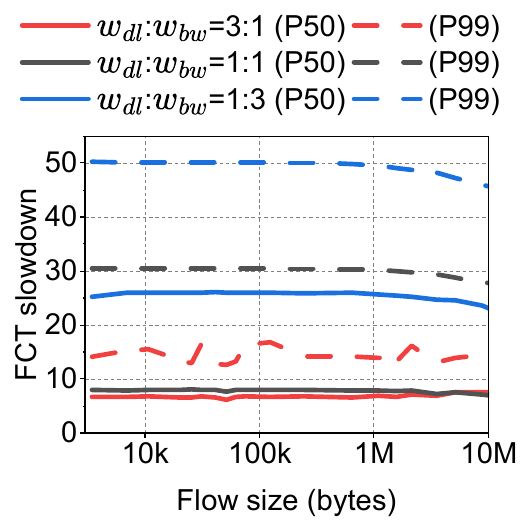} 
        \caption{\centering [Path-quality weights analysis] Weights tuples $(w_{dl},w_{lc})=(3,1),(1,1),(1,3)$}
        \label{fig:discuss-path-weights}
    \end{subfigure}
    \begin{subfigure}[T]{0.23\textwidth}  
        \includegraphics[width=\linewidth]{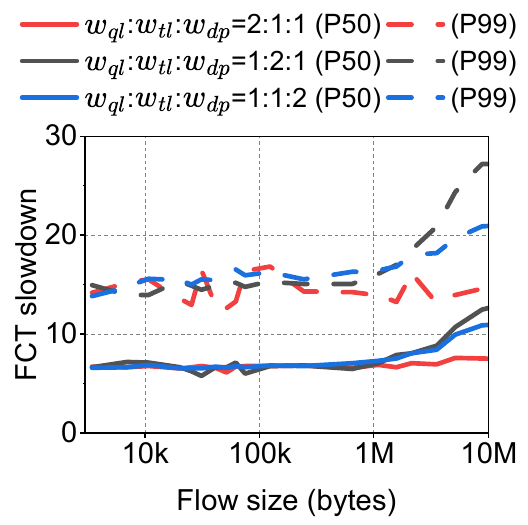} 
        \caption{\centering [Congestion-cost weights analysis] Weight tuples $(w_{ql},w_{tl},w_{dp})=(2,1,1),(1,2,1),(1,1,2)$.}
        \label{fig:discuss-cong-weights}
    \end{subfigure}
  \caption{[Sensitivity analysis] Median and tail FCT slowdown for WebSearch on the 8-DC topology at 30\% load.}
  \Description{[Sensitivity analysis] Median and tail FCT slowdown for WebSearch on the 8-DC topology at 30\% load.}
  \label{fig:discuss-total}
\end{figure*}

We present ablation and parameter-sensitivity results in this section. These experiments show how to configure \name and why each component matters in practice. The experiments measure the impact of the control-plane path-quality term and the data-plane congestion term. They also identify robust integer-weight defaults for heterogeneous inter-DC deployments. Unless noted otherwise, all runs use the Web Search workload at 30\% load using DCQCN as the default CC.

\subsection{Ablation Sensitivity Analysis}
\label{sec:discussion-ablation}
We run three variants on the 8-DC topology (\ref{subfig:moti-porCor}): 
\begin{itemize}[topsep=0pt]
    \item \texttt{rm-alpha} — path-quality removed ($\alpha\!=\!0$);
    \item \texttt{rm-beta} — congestion removed ($\beta\!=\!0$);
    \item full \name with representative $(\alpha,\beta)$ settings.
\end{itemize}

\paragraph{Key findings.}
\figref{fig:discuss-ablation} shows two clear failure modes. First, the \texttt{rm-alpha} run (path-quality removed) severely degrades performance across almost all flow sizes. For example, the median for a 3,438 B flow rises from 6.8 (normal) to 26.0 when $\alpha=0$ (+280\%). The P99 for the same size rises from 12.1 to 50.0 (+312\%). The \texttt{rm-alpha} curve stays well above the others for the entire flow-size range. This pattern means that using only on-switch congestion signals tends to place flows on high-delay routes in this heterogeneous topology. Second, the \texttt{rm-beta} run (congestion removed) preserves medians for small and mid-sized flows but fails for large transfers. For the largest flows (29.7 MB) the median increases from 8.7 (normal) to 31.2 (+260\%) and P99 jumps from 17.1 to 58.4 (+240\%). This shows that path-only selection cannot prevent contention among long-lived elephants. The full \name run consistently achieves the lowest and most stable p50 and P99 across sizes.

\paragraph{Takeaway.}
Both components are necessary. The control-plane path-quality term prevents systematic placement on high-delay links and thus keeps medians low. The on-switch congestion term prevents herd-driven contention among large flows and thus controls tails. In practice, operators should use a fused cost with non-zero $\alpha$ and $\beta$. A modest bias toward path quality (e.g., $\alpha=3,\beta=1$) yields a robust trade-off between median and tail in capacity–delay asymmetric inter-DC deployments.

\subsection{Global Fusion-Weight Sensitivity Analysis}
\label{sec:discussion-global-weights}
We sweep global fusion weights $(\alpha,\beta)\in\{(3,1)$, $(1,1)$, $(1,3)\}$ on the 8-DC topology.

\paragraph{Key findings.}
As shown in \figref{fig:discuss-global-weights}, all three weight settings produce similar medians. The delay-biased setting \((3,1)\) matches others on p50. The delay-biased setting, however, yields much smaller tails. Typical P99 values under \((3,1)\) fall in the 12–16 range. The balanced \((1,1)\) and congestion-biased \((1,3)\) settings show P99 values around 24–30 for many sizes. In short, prioritizing the control-plane path-quality term reduces P99 by roughly half compared to balanced or congestion-heavy choices, while leaving medians essentially unchanged.

\paragraph{Takeaway.}

When bandwidth and delay misaligned, favor path-quality in the fusion. A delay-biased fusion (e.g., $\alpha=3,\beta=1$) gives the most stable tails without hurting medians. Balanced or congestion-heavy weightings make the system more likely to over-react to transient signals and to send latency-sensitive flows onto high-capacity but slow links.

\subsection{Path-Quality Weight Sensitivity Analysis}
\label{sec:discussion-path-weights}
We vary $(w_{dl},w_{lc})\in\{(3,1)$, $(1,1)$, $(1,3)\}$ inside $C_{\mathrm{path}}$.

\paragraph{Key findings.}
As shown in \figref{fig:discuss-path-weights}, The delay-biased path score \((3,1)\) gives the best medians and tails. Under \((3,1)\) p50 values cluster near 6.1–7.6 and P99 near 12–17. The balanced \((1,1)\) choice yields slightly worse medians (7.0–8.1) and much larger tails (27–31). The capacity-biased \((1,3)\) choice performs worst: it raises medians and tails dramatically (p50 often > 20 and P99 in the 43–50 range for many sizes). Overall, weighting delay more than bandwidth halves P99 versus balanced settings and reduces medians by roughly 10–20\% compared to the balanced choice.

\paragraph{Takeaway.}
When capacity and latency trade off, give higher weight to delay in $C_{\mathrm{path}}$. A delay-biased setting (e.g., $w_{dl}$:$w_{lc}=3$:$1$) avoids placing latency-sensitive flows on high-capacity but slow links. This choice improves both median and tail FCT.

\subsection{Congestion-Cost Weight Sensitivity Analysis}
\label{sec:discussion-cong-weights}
We compare allocations $(w_{ql},w_{tl},w_{dp})$ $\in$ $\{(2,1,1)$, $(1,2,1)$, $(1,1,2)\}$ for $C_{\mathrm{cong}}$.

\paragraph{Key findings.}
In \figref{fig:discuss-cong-weights}, the three allocations show similar medians for small and mid flows. They diverge for large flows and in the tail. The queue-focused setting $(2,1,1)$ gives the most stable behavior: p50 stays near 6.1–7.6 and P99 near 12–17. The trend-heavy $(1,2,1)$ and duration-heavy $(1,1,2)$ settings raise P99 for the largest flows. These settings also increase p50 for the largest sizes (from $\approx$6–7 up to $\approx$ 8–14). The queue-focused choice keeps both medians and tails lower.

\paragraph{Takeaway.}
These results indicate that putting most weight on instantaneous queue level is the safest and most robust choice. A queue-first allocation (e.g., 2:1:1) limits P99 inflation while keeping medians stable. Emphasizing short-term trend or persistent-duration penalties can help very short flows but risks concentrating elephants onto fewer paths and amplifying noise. Therefore we recommend a conservative, queue-focused default (e.g., 2:1:1) for production deployments where path diversity and capacity–delay trade-offs exist.

\subsection{Limitations}
\label{sec:limitations}
While \name reduces placement inefficiencies caused by topology heterogeneity, it has two practical limitations that point to future work.

\textbf{Flow-level stickiness limits responsiveness.}
\name pins a flow to a chosen egress to preserve in-order delivery. 
\reviseadd{To B,C,D: CR-3}{We explicitly avoid migrating active flows (re-routing) because shifting paths mid-flow inevitably causes packet reordering, which triggers severe throughput collapse in RNICs due to Go-Back-N behavior. Instead, LCMP optimizes the initial placement to minimize collisions and delegates the handling of subsequent bursts to end-host CC. While this design prioritizes path consistency over mid-flow agility, it ensures correctness and stability on today's hardware.}

\textbf{RNIC out-of-order handling.}
The stickiness stems from RNICs' sensitivity to out-of-order (OoO) packets and their loss-recovery semantics. Many commodity RNICs treat OoO arrivals as losses and trigger retransmission. Aggressive per-packet or per-flowlet steering can therefore increase retransmits and hurt latency. Recent work shows promising directions to relax this constraint (e.g., in-network reordering and lightweight OoO tracking)\cite{mittal2018RevisitingNetwork,song2023ConWeave,wang2023SRNICScalable,huang2025FastScalable,li2025RevisitingRDMA,huang2024LEFTLightwEight}, but such techniques are not yet widely deployed.

\paragraph{Future directions.}
We highlight two practical research directions. First, explore fine-grained steering with OoO tolerance. We will combine selective per-flowlet or per-packet routing with lightweight in-network reordering or RNIC-side OoO tracking. The goal is to trade a small, controlled amount of reordering for much faster congestion reaction. Second, pursue cross-layer co-design with congestion control and loss recovery. We will align routing decisions with transport-layer signals so steering does not conflict with senders’ recovery logic.

\section{Related Work}
\label{sec:relatedWork}

\reviseadd{To A, E: CR-1.1}{
\paragraph{Comparison with WAN traffic engineering}
Recent WAN TE schemes like POP\cite{nara2021POP}, Teal\cite{xu2023TealLearningAccelerated} and RedTE\cite{gui2024RedTEMitigating} optimize global throughput by adjusting traffic splitting ratios. However, they operate on timescales (ms-level) that, while effective for TCP, are insufficient for RDMA. Long-haul RDMA requires µs-scale reaction to prevent PFC storms caused by transient microbursts. Furthermore, dynamic TE adjustments\cite{he2023rthop,diao2024dynlowlet,liu2024drlrouting} can introduce packet reordering. Unlike TCP, RNICs rely on Go-Back-N, where reordering triggers severe throughput collapse. LCMP complements WAN TE by performing fine-grained, reordering-free load balancing in the data plane at line rate to satisfy RDMA's strict latency and ordering constraints.
}

\paragraph{Long-haul link transport optimization}
The expansion of large-scale DCs is constrained by limited land and power resources. To overcome them, major cloud service providers (CSPs) deploy multiple DCs interconnected through dedicated optical fibers.
Recent efforts have focused on optimizing transport over long-haul networks. SWING\cite{chen2022swing} proposes a PFC relay mechanism that extends lossless RDMA to long-haul links.
Bifrost\cite{huang2024minimizing} introduces a downstream-driven lossless flow control to support cross-DC data transfers over long distances, achieving low buffer reservation, and zero packet loss. Considering the characteristics of long-haul links with large RTT and BDP, LSCC\cite{long2024lscc} proposes a link-segmented CC algorithm for inter-DC networks, which leverages more fine-grained control signals to achieve high throughput and low latency over long-haul links.

\paragraph{Inter-DC transport and routing optimization}
Inter-DC fabrics, with µs-scale RTTs and heterogeneous link capacities, have been addressed largely by two strands of work: control-plane traffic engineering and CC, but not by routing algorithm that jointly considers path quality and on-switch signals. Centralized TE\cite{jain2013B4Experience, singh2015JupiterRising, yap2017TakingEdge, ferguson2021OrionGoogles, zhang2018BDSCentralized, zhang2021BDSInterDatacenter} yields high steady-state utilization via global optimization yet acts at coarse timescales and cannot make per-flow packet-time choices to avoid short-lived tail spikes. Transport and hybrid proposals that fuse ECN, delay or in-band telemetry\cite{zeng2022CongestionControla, geng2023DelayBased} improve end-to-end rate control but generally leave path selection to ECMP. Recent systems\cite{long2024LSCCLinksegmented, li2025RevisitingRDMA, lv2025OmniDMAScalable} reduce feedback latency or strengthen transport semantics, yet they either require costly deployment changes or retain default multipath routing. In short, prior inter-DC work improves global planning or transport behavior but does not provide a distributed, data-plane feasible routing method.

\paragraph{Intra-DC routing, load balancing and CC}
Intra-DC routing and CC methods address lossless delivery, and reordering sensitivity, but existing schemes typically assume µs-scale feedback, or centralized coordination, which is incompatible with the long RTTs, path heterogeneity, and herd effects we identify in \circled{C1}–\circled{C3}.  Early multipath adaptations\cite{zhou2014WCMPWeighted, lu2018MultiPathTransport, alizadeh2014CONGADistributed, katta2016HULAScalable, ghorbani2017DRILLMicro, katta2017CloveCongestionaware, zhang2017ResilientDatacenterb, zhang2021HashingLinearity, wetherall2023ImprovingNetwork, liu2025UnlockingECMP, luo2025SeqBalanceCongestionaware} improve fairness or throughput via static weights or flow-splitting, yet they either lack realtime congestion awareness or risk RDMA-unfriendly reordering. RDMA congestion controllers and telemetry-driven designs\cite{zhu2015CongestionControl,li2019HPCCHigh,mittal2015TIMELYRTTbased, kumar2020SwiftDelay, saeed2020AnnulusDual, taheri2020RoCCRobust, addanki2022PowerTCPPushing, goyal2022BackpressureFlow, zhong2022PACCProactive, chen2023SwingProviding, zhang2023RCCEnabling, wu2024COERNetwork, zhang2024PACCProactive, zou2024AchievingUltralow, wan2025RHCCRevisiting, zhang2025MORS} provide valuable signals for rate control but leave routing to ECMP and their feedback is outdated across inter-DC RTTs.  Flowlet and sequencing approaches\cite{chen2024HF^2THostBaseda,besta2020FatPathsRouting,luo2025SeqBalanceCongestionaware} reduce reordering or enable finer steering but depend on host changes, or central schedulers, constraints that limit their usefulness.  Recent hardware efforts\cite{liu2025UnlockingECMP, li2024UCMP} advance switch-side steering but do not consider congestion signals and path quality. \name differs by preserving per-flow path consistency, fusing path quality with congestion estimates, and using a low-state selection.

\section{Conclusion}
\label{sec:conclusion}

We presented \name, a distributed long-haul cost-aware multi-path routing framework for inter-DC networks. \name fuses a path-quality score with on-switch congestion signals and applies a diversity-preserving selection step to make line-rate multi-path decisions.

Our evaluation on a small-scale testbed and large-scale NS-3 simulations under the \emph{2000 km inter-DC scenario} demonstrates that this design consistently improves flow-completion behavior and is robust across realistic workloads and CC algorithms.

We currently enforce per-flow stickiness to preserve RDMA in-order delivery, which limits aggressive rebalancing under sudden congestion. Future work will explore fine-grained steering with lightweight out-of-order tolerance and tighter routing–congestion-control co-design to restore responsiveness without sacrificing correctness.

\begin{acks}
We would like to thank our shepherd Yang Zhou and anonymous reviewers for their valuable and constructive feedback. This work is supported by the National Key R\&D Program of China under Grant 2024YFB2906900, the Beijing Nova Program under Grant 2023140, the Key Program of the Beijing Natural Science Foundation (Haidian Original Innovation Joint Fund) under Grant L252013, and the National Natural Science Foundation of China for Distinguished Young Scholars under Grant 62425201. 
\end{acks}

\bibliographystyle{ACM-Reference-Format}
\bibliography{reference}

@INPROCEEDINGS{zhang2025MORS,
  author={Zhang, Yuchao and Zheng, Chenyue and Wu, Wenfei and Jiang, Zhuo and Wang, Lei and Dai, Huichen and Zhang, Zhang and Nie, Jianglong and Wang, Wendong},
  booktitle={2025 IEEE 33rd International Conference on Network Protocols (ICNP)}, 
  title={MORS: Traffic-Aware Routing based on Temporal Attributes for Model Training Clusters}, 
  year={2025},
  volume={},
  number={},
  pages={1-12},
  }

@article{diao2024dynlowlet,
  title = {Deep Reinforcement Learning Based Dynamic Flowlet Switching for {{DCN}}},
  author = {Diao, Xinglong and Gu, Huaxi and Wei, Wenting and Jiang, Guoyong and Li, Baochun},
  year = 2024,
  month = apr,
  journal = {IEEE Transactions on Cloud Computing},
  volume = {12},
  number = {2},
  pages = {580--593},
  issn = {2168-7161}
}

@article{liu2024drlrouting,
  title = {Deep Distributional Reinforcement Learning-Based Adaptive Routing with Guaranteed Delay Bounds},
  author = {Liu, Jianmin and Li, Dan and Xu, Yongjun},
  year = 2024,
  month = dec,
  journal = {IEEE/ACM Transactions on Networking},
  volume = {32},
  number = {6},
  pages = {4692--4706},
  issn = {1558-2566},
  langid = {american}
}

@article{he2023rthop,
  title = {{{RTHop}}: {{Real-time}} Hop-by-Hop Mobile Network Routing by Decentralized Learning with Semantic Attention},
  author = {He, Bo and Wang, Jingyu and Qi, Qi and Sun, Haifeng and Liao, Jianxin},
  year = 2023,
  month = mar,
  journal = {IEEE Transactions on Mobile Computing},
  volume = {22},
  number = {3},
  pages = {1731--1747},
  issn = {1558-0660},
  langid = {american}
}

@inproceedings{nara2021POP,
author = {Narayanan, Deepak and Kazhamiaka, Fiodar and Abuzaid, Firas and Kraft, Peter and Agrawal, Akshay and Kandula, Srikanth and Boyd, Stephen and Zaharia, Matei},
title = {Solving Large-Scale Granular Resource Allocation Problems Efficiently with POP},
year = {2021},
isbn = {9781450387095},
publisher = {Association for Computing Machinery},
address = {New York, NY, USA},
booktitle = {Proceedings of the ACM SIGOPS 28th Symposium on Operating Systems Principles},
pages = {521–537},
numpages = {17},
location = {Virtual Event, Germany},
series = {SOSP '21}
}

@inproceedings{gui2024RedTEMitigating,
  title = {RedTE: Mitigating Subsecond Traffic Bursts with Real-Time and Distributed Traffic Engineering},
  shorttitle = {RedTE},
  booktitle = {Proceedings of the ACM SIGCOMM 2024 Conference},
  author = {Gui, Fei and Wang, Songtao and Li, Dan and Chen, Li and Gao, Kaihui and Min, Congcong and Wang, Yi},
  year = 2024,
  month = aug,
  pages = {71--85},
  publisher = {Association for Computing Machinery},
  address = {New York, NY, USA},
  isbn = {979-8-4007-0614-1}
}

@inproceedings{xu2023TealLearningAccelerated,
  title = {Teal: Learning-Accelerated Optimization of WAN Traffic Engineering},
  shorttitle = {Teal},
  booktitle = {Proceedings of the ACM SIGCOMM 2023 Conference},
  author = {Xu, Zhiying and Yan, Francis Y. and Singh, Rachee and Chiu, Justin T. and Rush, Alexander M. and Yu, Minlan},
  year = 2023,
  month = sep,
  pages = {378--393},
  publisher = {ACM},
  address = {New York NY USA},
  isbn = {979-8-4007-0236-5}
}

@inproceedings{song2023ConWeave,
  title = {Network Load Balancing with In-Network Reordering Support for RDMA},
  booktitle = {Proceedings of the ACM SIGCOMM 2023 Conference},
  author = {Song, Cha Hwan and Khooi, Xin Zhe and Joshi, Raj and Choi, Inho and Li, Jialin and Chan, Mun Choon},
  year = {2023},
  pages = {816--831},
  publisher = {Association for Computing Machinery},
  address = {New York, NY, USA},
  isbn = {979-8-4007-0236-5}
}

@inproceedings{alfaresScalableCommodity2008,
  title = {A Scalable, Commodity Data Center Network Architecture},
  booktitle = {Proceedings of the ACM SIGCOMM 2008 Conference on Data Communication},
  author = {{Al-Fares}, Mohammad and Loukissas, Alexander and Vahdat, Amin},
  year = {2008},
  pages = {63--74},
  publisher = {Association for Computing Machinery},
  address = {New York, NY, USA},
  isbn = {978-1-60558-175-0}
}

@INPROCEEDINGS{huang2025FastScalable,
  author={Huang, Peihao and Chen, Guo and Zhang, Xin and Liu, Can and Wang, Hongyu and Shen, Huijun and Bian, Ying and Lu, Yuanwei and Ruan, Zhenyuan and Li, Bojie and Zhang, Jiansong and Liu, Yongfeng and Chen, Zhigang},
  booktitle={IEEE INFOCOM 2025 - IEEE Conference on Computer Communications}, 
  title={Fast and Scalable Selective Retransmission for RDMA}, 
  year={2025},
  volume={},
  number={},
  pages={1-10},
}

@inproceedings{roy2015SocialNetwork,
  title = {Inside the Social Network's (Datacenter) Network},
  booktitle = {Proceedings of the 2015 ACM Conference on Special Interest Group on Data Communication},
  author = {Roy, Arjun and Zeng, Hongyi and Bagga, Jasmeet and Porter, George and Snoeren, Alex C.},
  year = {2015},
  pages = {123--137},
  publisher = {Association for Computing Machinery},
  address = {New York, NY, USA},
  isbn = {978-1-4503-3542-3}
}

@article{alizadeh2010DCTCP,
  title = {Data Center TCP (DCTCP)},
  author = {Alizadeh, Mohammad and {View Profile} and Greenberg, Albert and {View Profile} and Maltz, David A. and {View Profile} and Padhye, Jitendra and {View Profile} and Patel, Parveen and {View Profile} and Prabhakar, Balaji and {View Profile} and Sengupta, Sudipta and {View Profile} and Sridharan, Murari and {View Profile}},
  year = {2010},
  month = aug,
  journal = {Proceedings of the ACM SIGCOMM 2010 conference},
  pages = {63--74},
  issn = {9781450302012},
  volume = {40},
  number = {4},
}

@inproceedings{li2024UCMP,
  title = {Uniform-Cost Multi-Path Routing for Reconfigurable Data Center Networks},
  booktitle = {Proceedings of the ACM SIGCOMM 2024 Conference},
  author = {Li, Jialong and Gong, Haotian and De Marchi, Federico and Gong, Aoyu and Lei, Yiming and Bai, Wei and Xia, Yiting},
  year = {2024},
  pages = {433--448},
  publisher = {Association for Computing Machinery},
  address = {New York, NY, USA},
  isbn = {979-8-4007-0614-1}
}

@inproceedings{geng2023DelayBased,
  title = {Delay Based Congestion Control for Cross-Datacenter Networks},
  booktitle = {2023 IEEE/ACM 31st International Symposium on Quality of Service (IWQoS)},
  author = {Geng, Yantao and Zhang, Han and Shi, Xingang and Wang, Jilong and Yin, Xia and He, Dongbiao and Li, Yahui},
  year = {2023},
  pages = {1--4},
}

@inproceedings{jain2013B4Experience,
  title = {B4: Experience with a Globally-Deployed Software Defined Wan},
  booktitle = {Proceedings of the ACM SIGCOMM 2013 Conference on SIGCOMM},
  author = {Jain, Sushant and Kumar, Alok and Mandal, Subhasree and Ong, Joon and Poutievski, Leon and Singh, Arjun and Venkata, Subbaiah and Wanderer, Jim and Zhou, Junlan and Zhu, Min and Zolla, Jon and H{\"o}lzle, Urs and Stuart, Stephen and Vahdat, Amin},
  year = {2013},
  pages = {3--14},
  publisher = {Association for Computing Machinery},
  address = {Hong Kong, China and New York, NY, USA},
  isbn = {978-1-4503-2056-6}
}

@inproceedings{li2025RevisitingRDMA,
  title = {Revisiting RDMA Reliability for Lossy Fabrics},
  booktitle = {Proceedings of the ACM SIGCOMM 2025 Conference},
  author = {Li, Wenxue and Liu, Xiangzhou and Zhang, Yunxuan and Wang, Zihao and Gu, Wei and Qian, Tao and Zeng, Gaoxiong and Ren, Shoushou and Huang, Xinyang and Ren, Zhenghang and Liu, Bowen and Zhang, Junxue and Chen, Kai and Liu, Bingyang},
  year = {2025},
  pages = {85--98},
  publisher = {Association for Computing Machinery},
  address = {New York, NY, USA},
  isbn = {979-8-4007-1524-2}
}

@inproceedings{long2024LSCCLinksegmented,
  title = {LSCC: Link-Segmented Congestion Control for RDMA in Cross-Datacenter Networks},
  booktitle = {2024 IEEE/ACM 32nd International Symposium on Quality of Service (IWQoS)},
  author = {Long, Minfei and Han, Jiangping and Wang, Wentao and Yang, Jiayu and Xue, Kaiping},
  year = {2024},
  pages = {1--10},
}

@inproceedings{lv2025OmniDMAScalable,
  title = {OmniDMA: Scalable RDMA Transport over WAN},
  booktitle = {Proceedings of the 9th Asia-Pacific Workshop on Networking},
  author = {Lv, Kai and Li, Jinyang and Zhang, Pengyi and Pan, Heng and Li, Luyang and Hu, Shuihai and Li, Zhenyu and Xie, Gaogang and Zhou, Jingbin and Tan, Kun},
  year = {2025},
  pages = {135--141},
  publisher = {Association for Computing Machinery},
  address = {New York, NY, USA},
  isbn = {979-8-4007-1401-6}
}

@article{zeng2022CongestionControla,
  title = {Congestion Control for Cross-Datacenter Networks},
  author = {Zeng, Gaoxiong and Bai, Wei and Chen, Ge and Chen, Kai and Han, Dongsu and Zhu, Yibo and Cui, Lei},
  year = {2022},
  journal = {IEEE/ACM Transactions on Networking},
  volume = {30},
  number = {5},
  pages = {2074--2089},
}

@inproceedings{besta2020FatPathsRouting,
  title = {FatPaths: Routing in Supercomputers and Data Centers When Shortest Paths Fall Short},
  booktitle = {SC20: International Conference for High Performance Computing, Networking, Storage and Analysis},
  author = {Besta, Maciej and Schneider, Marcel and Konieczny, Marek and Cynk, Karolina and Henriksson, Erik and Girolamo, Salvatore Di and Singla, Ankit and Hoefler, Torsten},
  year = {2020},
  pages = {1--18},
}

@inproceedings{chen2024HF^2THostBaseda,
  title = {HF{\textasciicircum}2T: Host-Based Flowlet Fine-Tuning for RDMA Load Balancing},
  shorttitle = {HF{\textasciicircum}2T},
  booktitle = {Proceedings of the 8th Asia-Pacific Workshop on Networking},
  author = {Chen, Chuhao and Ye, Jiarui and Gao, Yongbo and Liu, Sen and Xu, Yang},
  year = {2024},
  month = aug,
  pages = {9--15},
  publisher = {ACM},
  address = {Sydney Australia},
  isbn = {979-8-4007-1758-1}
}

@inproceedings{li2019HPCCHigh,
  title = {HPCC: High Precision Congestion Control},
  shorttitle = {HPCC},
  booktitle = {Proceedings of the ACM Special Interest Group on Data Communication},
  author = {Li, Yuliang and Miao, Rui and Liu, Hongqiang Harry and Zhuang, Yan and Feng, Fei and Tang, Lingbo and Cao, Zheng and Zhang, Ming and Kelly, Frank and Alizadeh, Mohammad and Yu, Minlan},
  year = {2019},
  month = aug,
  pages = {44--58},
  publisher = {ACM},
  address = {Beijing China},
  isbn = {978-1-4503-5956-6}
}

@inproceedings{liu2025UnlockingECMP,
  title = {Unlocking ECMP Programmability for Precise Traffic Control},
  booktitle = {22nd USENIX Symposium on Networked Systems Design and Implementation (NSDI 25)},
  author = {Liu, Yadong and Xiao, Yunming and Zhang, Xuan and Dang, Weizhen and Liu, Huihui and Li, Xiang and He, Zekun and Wang, Jilong and Kuzmanovic, Aleksandar and Chen, Ang and Miao, Congcong},
  year = {2025},
  month = apr,
  pages = {87--106},
  publisher = {USENIX Association},
  address = {Philadelphia, PA},
  isbn = {978-1-939133-46-5}
}

@inproceedings{lu2018MultiPathTransport,
  title = {Multi-Path Transport for RDMA in Datacenters},
  booktitle = {15th USENIX Symposium on Networked Systems Design and Implementation (NSDI 18)},
  author = {Lu, Yuanwei and Chen, Guo and Li, Bojie and Tan, Kun and Xiong, Yongqiang and Cheng, Peng and Zhang, Jiansong and Chen, Enhong and Moscibroda, Thomas},
  year = {2018},
  month = apr,
  pages = {357--371},
  publisher = {USENIX Association},
  address = {Renton, WA},
  isbn = {978-1-939133-01-4}
}

@article{luo2025SeqBalanceCongestionaware,
  title = {SeqBalance: Congestion-Aware Load Balancing with No Reordering in Data Center Networks},
  author = {Luo, Huimin and Zhang, Jiao and Yu, Mingxuan and Pan, Yongchen and Pan, Tian and Huang, Tao},
  year = {2025},
  journal = {IEEE Internet of Things Journal},
  volume = {12},
  number = {13},
  pages = {25707--25719},
}

@article{mittal2015TIMELYRTTbased,
  title = {TIMELY: RTT-Based Congestion Control for the Datacenter},
  shorttitle = {TIMELY},
  author = {Mittal, Radhika and Lam, Vinh The and Dukkipati, Nandita and Blem, Emily and Wassel, Hassan and Ghobadi, Monia and Vahdat, Amin and Wang, Yaogong and Wetherall, David and Zats, David},
  year = {2015},
  month = sep,
  journal = {ACM SIGCOMM Computer Communication Review},
  volume = {45},
  number = {4},
  pages = {537--550},
  issn = {0146-4833},
}

@inproceedings{zhou2014WCMPWeighted,
  title = {WCMP: Weighted Cost Multipathing for Improved Fairness in Data Centers},
  booktitle = {Proceedings of the Ninth European Conference on Computer Systems},
  author = {Zhou, Junlan and Tewari, Malveeka and Zhu, Min and Kabbani, Abdul and Poutievski, Leon and Singh, Arjun and Vahdat, Amin},
  year = {2014},
  publisher = {Association for Computing Machinery},
  address = {New York, NY, USA},
  isbn = {978-1-4503-2704-6},
  numpages = {14}
}

@inproceedings{zhu2015CongestionControl,
  title = {Congestion Control for Large-Scale RDMA Deployments},
  booktitle = {Proceedings of the 2015 ACM Conference on Special Interest Group on Data Communication},
  author = {Zhu, Yibo and Eran, Haggai and Firestone, Daniel and Guo, Chuanxiong and Lipshteyn, Marina and Liron, Yehonatan and Padhye, Jitendra and Raindel, Shachar and Yahia, Mohamad Haj and Zhang, Ming},
  year = {2015},
  month = sep,
  volume = {45},
  pages = {523--536},
  publisher = {Association for Computing Machinery},
  address = {New York, NY, USA},
}

@inproceedings{bai2023EmpoweringAzure,
  title = {Empowering Azure Storage with RDMA},
  booktitle = {20th USENIX Symposium on Networked Systems Design and Implementation (NSDI 23)},
  author = {Bai, Wei and Abdeen, Shanim Sainul and Agrawal, Ankit and Attre, Krishan Kumar and Bahl, Paramvir and Bhagat, Ameya and Bhaskara, Gowri and Brokhman, Tanya and Cao, Lei and Cheema, Ahmad and Chow, Rebecca and Cohen, Jeff and Elhaddad, Mahmoud and Ette, Vivek and Figlin, Igal and Firestone, Daniel and George, Mathew and German, Ilya and Ghai, Lakhmeet and Green, Eric and Greenberg, Albert and Gupta, Manish and Haagens, Randy and Hendel, Matthew and Howlader, Ridwan and John, Neetha and Johnstone, Julia and Jolly, Tom and Kramer, Greg and Kruse, David and Kumar, Ankit and Lan, Erica and Lee, Ivan and Levy, Avi and Lipshteyn, Marina and Liu, Xin and Liu, Chen and Lu, Guohan and Lu, Yuemin and Lu, Xiakun and Makhervaks, Vadim and Malashanka, Ulad and Maltz, David A. and Marinos, Ilias and Mehta, Rohan and Murthi, Sharda and Namdhari, Anup and Ogus, Aaron and Padhye, Jitendra and Pandya, Madhav and Phillips, Douglas and Power, Adrian and Puri, Suraj and Raindel, Shachar and Rhee, Jordan and Russo, Anthony and Sah, Maneesh and Sheriff, Ali and Sparacino, Chris and Srivastava, Ashutosh and Sun, Weixiang and Swanson, Nick and Tian, Fuhou and Tomczyk, Lukasz and Vadlamuri, Vamsi and Wolman, Alec and Xie, Ying and Yom, Joyce and Yuan, Lihua and Zhang, Yanzhao and Zill, Brian},
  year = {2023},
  month = apr,
  pages = {49--67},
  publisher = {USENIX Association},
  address = {Boston, MA},
  isbn = {978-1-939133-33-5}
}

@inproceedings{gangidi2024RDMAEthernet,
  title = {RDMA over Ethernet for Distributed Training at Meta Scale},
  booktitle = {Proceedings of the ACM SIGCOMM 2024 Conference},
  author = {Gangidi, Adithya and Miao, Rui and Zheng, Shengbao and Bondu, Sai Jayesh and Goes, Guilherme and Morsy, Hany and Puri, Rohit and Riftadi, Mohammad and Shetty, Ashmitha Jeevaraj and Yang, Jingyi and Zhang, Shuqiang and Fernandez, Mikel Jimenez and Gandham, Shashidhar and Zeng, Hongyi},
  year = {2024},
  pages = {57--70},
  publisher = {Association for Computing Machinery},
  address = {New York, NY, USA},
  isbn = {979-8-4007-0614-1}
}

@inproceedings{gao2021WhenCloud,
  title = {When Cloud Storage Meets RDMA},
  booktitle = {18th USENIX Symposium on Networked Systems Design and Implementation (NSDI 21)},
  author = {Gao, Yixiao and Li, Qiang and Tang, Lingbo and Xi, Yongqing and Zhang, Pengcheng and Peng, Wenwen and Li, Bo and Wu, Yaohui and Liu, Shaozong and Yan, Lei and Feng, Fei and Zhuang, Yan and Liu, Fan and Liu, Pan and Liu, Xingkui and Wu, Zhongjie and Wu, Junping and Cao, Zheng and Tian, Chen and Wu, Jinbo and Zhu, Jiaji and Wang, Haiyong and Cai, Dennis and Wu, Jiesheng},
  year = {2021},
  month = apr,
  pages = {519--533},
  publisher = {USENIX Association},
  isbn = {978-1-939133-21-2},
}

@inproceedings{ferguson2021OrionGoogles,
  title = {Orion: Google's Software-Defined Networking Control Plane},
  booktitle = {18th USENIX Symposium on Networked Systems Design and Implementation (NSDI 21)},
  author = {Ferguson, Andrew D. and Gribble, Steve and Hong, Chi-Yao and Killian, Charles and Mohsin, Waqar and Muehe, Henrik and Ong, Joon and Poutievski, Leon and Singh, Arjun and Vicisano, Lorenzo and Alimi, Richard and Chen, Shawn Shuoshuo and Conley, Mike and Mandal, Subhasree and Nagaraj, Karthik and Bollineni, Kondapa Naidu and Sabaa, Amr and Zhang, Shidong and Zhu, Min and Vahdat, Amin},
  year = {2021},
  month = apr,
  pages = {83--98},
  publisher = {USENIX Association},
  isbn = {978-1-939133-21-2}
}

@inproceedings{singh2015JupiterRising,
  title = {Jupiter Rising: A Decade of Clos Topologies and Centralized Control in Google's Datacenter Network},
  booktitle = {Proceedings of the 2015 ACM Conference on Special Interest Group on Data Communication},
  author = {Singh, Arjun and Ong, Joon and Agarwal, Amit and Anderson, Glen and Armistead, Ashby and Bannon, Roy and Boving, Seb and Desai, Gaurav and Felderman, Bob and Germano, Paulie and Kanagala, Anand and Provost, Jeff and Simmons, Jason and Tanda, Eiichi and Wanderer, Jim and H{\"o}lzle, Urs and Stuart, Stephen and Vahdat, Amin},
  year = {2015},
  pages = {183--197},
  publisher = {Association for Computing Machinery},
  address = {London, United Kingdom and New York, NY, USA},
  isbn = {978-1-4503-3542-3}
}

@inproceedings{yap2017TakingEdge,
  title = {Taking the Edge off with Espresso: Scale, Reliability and Programmability for Global Internet Peering},
  booktitle = {Proceedings of the Conference of the ACM Special Interest Group on Data Communication},
  author = {Yap, Kok-Kiong and Motiwala, Murtaza and Rahe, Jeremy and Padgett, Steve and Holliman, Matthew and Baldus, Gary and Hines, Marcus and Kim, Taeeun and Narayanan, Ashok and Jain, Ankur and Lin, Victor and Rice, Colin and Rogan, Brian and Singh, Arjun and Tanaka, Bert and Verma, Manish and Sood, Puneet and Tariq, Mukarram and Tierney, Matt and Trumic, Dzevad and Valancius, Vytautas and Ying, Calvin and Kallahalla, Mahesh and Koley, Bikash and Vahdat, Amin},
  year = {2017},
  pages = {432--445},
  publisher = {Association for Computing Machinery},
  address = {Los Angeles, CA, USA and New York, NY, USA},
  isbn = {978-1-4503-4653-5}
}

@inproceedings{mittal2018RevisitingNetwork,
  title = {Revisiting Network Support for RDMA},
  booktitle = {Proceedings of the 2018 Conference of the ACM Special Interest Group on Data Communication},
  author = {Mittal, Radhika and Shpiner, Alexander and Panda, Aurojit and Zahavi, Eitan and Krishnamurthy, Arvind and Ratnasamy, Sylvia and Shenker, Scott},
  year = {2018},
  month = aug,
  pages = {313--326},
  publisher = {Association for Computing Machinery},
  address = {New York, NY, USA},
  isbn = {978-1-4503-5567-4}
}

@inproceedings{wang2023SRNICScalable,
  title = {SRNIC: A Scalable Architecture for RDMA NICs},
  booktitle = {20th USENIX Symposium on Networked Systems Design and Implementation},
  author = {Wang, Zilong and Luo, Layong and Ning, Qingsong and Zeng, Chaoliang and Li, Wenxue and Wan, Xinchen and Xie, Peng and Feng, Tao and Cheng, Ke and Geng, Xiongfei and Wang, Tianhao and Ling, Weicheng and Huo, Kejia and An, Pingbo and Ji, Kui and Zhang, Shideng and Xu, Bin and Feng, Ruiqing and Ding, Tao and Chen, Kai and Guo, Chuanxiong},
  year = {2023},
  month = apr,
  pages = {1--14},
  publisher = {USENIX Association},
  address = {Boston, MA},
  isbn = {978-1-939133-33-5}
}

@inproceedings{huang2024LEFTLightwEight,
  title = {LEFT: LightwEight and FasT Packet Reordering for RDMA},
  shorttitle = {LEFT},
  booktitle = {Proceedings of the 8th Asia-Pacific Workshop on Networking},
  author = {Huang, Peihao and Zhang, Xin and Chen, Zhigang and Liu, Can and Chen, Guo},
  year = {2024},
  month = aug,
  pages = {67--73},
  publisher = {Association for Computing Machinery},
  address = {New York, NY, USA},
  isbn = {979-8-4007-1758-1}
}

@inproceedings{addanki2022PowerTCPPushing,
  title = {PowerTCP: Pushing the Performance Limits of Datacenter Networks},
  booktitle = {19th USENIX Symposium on Networked Systems Design and Implementation (NSDI 22)},
  author = {Addanki, Vamsi and Michel, Oliver and Schmid, Stefan},
  year = {2022},
  month = apr,
  pages = {51--70},
  publisher = {USENIX Association},
  address = {Renton, WA},
  isbn = {978-1-939133-27-4}
}

@article{chen2023SwingProviding,
  title = {Swing: Providing Long-Range Lossless RDMA via PFC-Relay},
  shorttitle = {Swing},
  author = {Chen, Yanqing and Tian, Chen and Dong, Jiaqing and Feng, Song and Zhang, Xu and Liu, Chang and Yu, Peiwen and Xia, Nai and Dou, Wanchun and Chen, Guihai},
  year = {2023},
  month = jan,
  journal = {IEEE Transactions on Parallel and Distributed Systems},
  volume = {34},
  number = {1},
  pages = {63--75},
  issn = {1558-2183},
}

@inproceedings{goyal2022BackpressureFlow,
  title = {Backpressure Flow Control},
  booktitle = {19th USENIX Symposium on Networked Systems Design and Implementation (NSDI 22)},
  author = {Goyal, Prateesh and Shah, Preey and Zhao, Kevin and Nikolaidis, Georgios and Alizadeh, Mohammad and Anderson, Thomas E.},
  year = {2022},
  month = apr,
  pages = {779--805},
  publisher = {USENIX Association},
  address = {Renton, WA},
  isbn = {978-1-939133-27-4}
}

@inproceedings{kumar2020SwiftDelay,
  title = {Swift: Delay Is Simple and Effective for Congestion Control in the Datacenter},
  shorttitle = {Swift},
  booktitle = {Proceedings of the Annual Conference of the ACM Special Interest Group on Data Communication on the Applications, Technologies, Architectures, and Protocols for Computer Communication},
  author = {Kumar, Gautam and Dukkipati, Nandita and Jang, Keon and Wassel, Hassan M. G. and Wu, Xian and Montazeri, Behnam and Wang, Yaogong and Springborn, Kevin and Alfeld, Christopher and Ryan, Michael and Wetherall, David and Vahdat, Amin},
  year = {2020},
  month = jul,
  pages = {514--528},
  publisher = {ACM},
  address = {Virtual Event USA},
  isbn = {978-1-4503-7955-7}
}

@inproceedings{saeed2020AnnulusDual,
  title = {Annulus: A Dual Congestion Control Loop for Datacenter and WAN Traffic Aggregates},
  shorttitle = {Annulus},
  booktitle = {Proceedings of the Annual Conference of the ACM Special Interest Group on Data Communication on the Applications, Technologies, Architectures, and Protocols for Computer Communication},
  author = {Saeed, Ahmed and Gupta, Varun and Goyal, Prateesh and Sharif, Milad and Pan, Rong and Ammar, Mostafa and Zegura, Ellen and Jang, Keon and Alizadeh, Mohammad and Kabbani, Abdul and Vahdat, Amin},
  year = {2020},
  month = jul,
  pages = {735--749},
  publisher = {ACM},
  address = {Virtual Event USA},
  isbn = {978-1-4503-7955-7}
}

@inproceedings{taheri2020RoCCRobust,
  title = {RoCC: Robust Congestion Control for RDMA},
  shorttitle = {RoCC},
  booktitle = {Proceedings of the 16th International Conference on Emerging Networking EXperiments and Technologies},
  author = {Taheri, Parvin and Menikkumbura, Danushka and Vanini, Erico and Fahmy, Sonia and Eugster, Patrick and Edsall, Tom},
  year = {2020},
  month = nov,
  pages = {17--30},
  publisher = {ACM},
  address = {Barcelona Spain},
  isbn = {978-1-4503-7948-9}
}

@article{wan2025RHCCRevisiting,
  title = {RHCC: Revisiting Intra-Host Congestion Control in RDMA Networks},
  shorttitle = {RHCC},
  author = {Wan, Zirui and Zhang, Jiao and Wang, Yuxiang and Liu, Kefei and Pan, Haoyu and Pan, Yongchen and Huang, Tao},
  year = {2025},
  journal = {IEEE Transactions on Networking},
  volume={33},
  number={3},
  pages = {1--14},
  issn = {2998-4157},
}

@article{wu2024COERNetwork,
  title = {COER: A Network Interface Offloading Architecture for RDMA and Congestion Control Protocol Codesign},
  shorttitle = {COER},
  author = {Wu, Ke and Dong, Dezun and Xu, Weixia},
  year = {2024},
  month = sep,
  journal = {ACM Transactions on Architecture and Code Optimization},
  volume = {21},
  number = {3},
  pages = {49:1--49:26},
  issn = {1544-3566},
}

@article{zhang2023RCCEnabling,
  title = {RCC: Enabling Receiver-Driven RDMA Congestion Control With Congestion Divide-and-Conquer in Datacenter Networks},
  shorttitle = {RCC},
  author = {Zhang, Jiao and Zhong, Xiaolong and Wan, Zirui and Tian, Yu and Pan, Tian and Huang, Tao},
  year = {2023},
  month = feb,
  journal = {IEEE/ACM Transactions on Networking},
  volume = {31},
  number = {1},
  pages = {103--117},
  issn = {1558-2566},
}

@article{zhang2024PACCProactive,
  title = {PACC: A Proactive CNP Generation Scheme for Datacenter Networks},
  shorttitle = {PACC},
  author = {Zhang, Jiao and Wang, Yuqing and Zhong, Xiaolong and Yu, Mingxuan and Pan, Haoyu and Zhang, Yali and Guan, Zixuan and Che, Biyao and Wan, Zirui and Pan, Tian and Huang, Tao},
  year = {2024},
  month = jun,
  journal = {IEEE/ACM Transactions on Networking},
  volume = {32},
  number = {3},
  pages = {2586--2599},
  issn = {1558-2566},
}

@inproceedings{zhong2022PACCProactive,
  title = {PACC: Proactive and Accurate Congestion Feedback for RDMA Congestion Control},
  shorttitle = {PACC},
  booktitle = {IEEE INFOCOM 2022 - IEEE Conference on Computer Communications},
  author = {Zhong, Xiaolong and Zhang, Jiao and Zhang, Yali and Guan, Zixuan and Wan, Zirui},
  year = {2022},
  month = may,
  pages = {2228--2237},
  issn = {2641-9874},
}

@inproceedings{zou2024AchievingUltralow,
  title = {Achieving Ultra-Low Latency for Timeout-Less Congestion Control in Data Center Networks},
  booktitle = {2024 IEEE International Symposium on Parallel and Distributed Processing with Applications (ISPA)},
  author = {Zou, Shaojun and Jiang, Yi and Qu, Jiacheng and Zhang, Tao and Hu, Yuanzhen and Peng, Yujie},
  year = {2024},
  month = oct,
  pages = {1439--1444},
  issn = {2158-9208},
}

@inproceedings{alizadeh2014CONGADistributed,
  title = {CONGA: Distributed Congestion-Aware Load Balancing for Datacenters},
  booktitle = {Proceedings of the 2014 ACM Conference on SIGCOMM},
  author = {Alizadeh, Mohammad and Edsall, Tom and Dharmapurikar, Sarang and Vaidyanathan, Ramanan and Chu, Kevin and Fingerhut, Andy and Lam, Vinh The and Matus, Francis and Pan, Rong and Yadav, Navindra and Varghese, George},
  year = {2014},
  pages = {503--514},
  publisher = {Association for Computing Machinery},
  address = {New York, NY, USA},
  isbn = {978-1-4503-2836-4}
}

@inproceedings{ghorbani2017DRILLMicro,
  title = {DRILL: Micro Load Balancing for Low-Latency Data Center Networks},
  booktitle = {Proceedings of the Conference of the ACM Special Interest Group on Data Communication},
  author = {Ghorbani, Soudeh and Yang, Zibin and Godfrey, P. Brighten and Ganjali, Yashar and Firoozshahian, Amin},
  year = {2017},
  pages = {225--238},
  publisher = {Association for Computing Machinery},
  address = {Los Angeles, CA, USA and New York, NY, USA},
  isbn = {978-1-4503-4653-5}
}

@inproceedings{katta2016HULAScalable,
  title = {HULA: Scalable Load Balancing Using Programmable Data Planes},
  booktitle = {Proceedings of the Symposium on SDN Research},
  author = {Katta, Naga and Hira, Mukesh and Kim, Changhoon and Sivaraman, Anirudh and Rexford, Jennifer},
  year = {2016},
  publisher = {Association for Computing Machinery},
  address = {New York, NY, USA},
  articleno = {10},
  isbn = {978-1-4503-4211-7},
  numpages = {12}
}

@inproceedings{katta2017CloveCongestionaware,
  title = {Clove: Congestion-Aware Load Balancing at the Virtual Edge},
  booktitle = {Proceedings of the 13th International Conference on Emerging Networking Experiments and Technologies},
  author = {Katta, Naga and Ghag, Aditi and Hira, Mukesh and Keslassy, Isaac and Bergman, Aran and Kim, Changhoon and Rexford, Jennifer},
  year = {2017},
  pages = {323--335},
  publisher = {Association for Computing Machinery},
  address = {Incheon, Republic of Korea and New York, NY, USA},
  isbn = {978-1-4503-5422-6}
}

@inproceedings{wetherall2023ImprovingNetwork,
  title = {Improving Network Availability with Protective ReRoute},
  booktitle = {Proceedings of the ACM SIGCOMM 2023 Conference},
  author = {Wetherall, David and Kabbani, Abdul and Jacobson, Van and Winget, Jim and Cheng, Yuchung and Morrey III, Charles B. and Moravapalle, Uma and Gill, Phillipa and Knight, Steven and Vahdat, Amin},
  year = {2023},
  pages = {684--695},
  publisher = {Association for Computing Machinery},
  address = {New York, NY, USA and New York, NY, USA},
  isbn = {979-8-4007-0236-5}
}

@inproceedings{zhang2017ResilientDatacenterb,
  title = {Resilient Datacenter Load Balancing in the Wild},
  booktitle = {Proceedings of the Conference of the ACM Special Interest Group on Data Communication},
  author = {Zhang, Hong and Zhang, Junxue and Bai, Wei and Chen, Kai and Chowdhury, Mosharaf},
  year = {2017},
  pages = {253--266},
  publisher = {Association for Computing Machinery},
  address = {Los Angeles, CA, USA and New York, NY, USA},
  isbn = {978-1-4503-4653-5}
}

@inproceedings{zhang2021HashingLinearity,
  title = {Hashing Linearity Enables Relative Path Control in Data Centers},
  booktitle = {2021 USENIX Annual Technical Conference (USENIX ATC 21)},
  author = {Zhang, Zhehui and Zheng, Haiyang and Hu, Jiayao and Yu, Xiangning and Qi, Chenchen and Shi, Xuemei and Wang, Guohui},
  year = {2021},
  month = jul,
  pages = {855--862},
  publisher = {USENIX Association},
  isbn = {978-1-939133-23-6}
}

@misc{hopps2000analysis,
    series =    {Request for Comments},
    number =    2992,
    howpublished =  {RFC 2992},
    publisher = {RFC Editor},
    author =    {Christian Hopps},
    title =     {{Analysis of an Equal-Cost Multi-Path Algorithm}},
    pagetotal = 8,
    year =      2000,
    month =     nov,
}

@article{knight2011TopologyZoo,
  title = {The Internet Topology Zoo},
  author = {Knight, Simon and Nguyen, Hung X. and Falkner, Nickolas and Bowden, Rhys and Roughan, Matthew},
  year = {2011},
  journal = {IEEE Journal on Selected Areas in Communications},
  volume = {29},
  number = {9},
  pages = {1765--1775},
}

@inproceedings{zhang2024FedRDMACommunicationefficient,
  title = {FedRDMA: Communication-Efficient Cross-Silo Federated LLM via Chunked RDMA Transmission},
  booktitle = {Proceedings of the 4th Workshop on Machine Learning and Systems},
  author = {Zhang, Zeling and Cai, Dongqi and Zhang, Yiran and Xu, Mengwei and Wang, Shangguang and Zhou, Ao},
  year = {2024},
  pages = {126--133},
  publisher = {Association for Computing Machinery},
  address = {New York, NY, USA},
  isbn = {979-8-4007-0541-0}
}

@inproceedings{chen2024PreciseData,
  title = {Precise Data Center Traffic Engineering with Constrained Hardware Resources},
  booktitle = {21st USENIX Symposium on Networked Systems Design and Implementation (NSDI 24)},
  author = {Chen, Shawn Shuoshuo and He, Keqiang and Wang, Rui and Seshan, Srinivasan and Steenkiste, Peter},
  year = {2024},
  month = apr,
  pages = {669--690},
  publisher = {USENIX Association},
  address = {Santa Clara, CA},
  isbn = {978-1-939133-39-7}
}

@inproceedings{guo2016RDMACommodity,
author = {Guo, Chuanxiong and Wu, Haitao and Deng, Zhong and Soni, Gaurav and Ye, Jianxi and Padhye, Jitu and Lipshteyn, Marina},
title = {RDMA over Commodity Ethernet at Scale},
year = {2016},
isbn = {9781450341936},
publisher = {Association for Computing Machinery},
address = {New York, NY, USA},
booktitle = {Proceedings of the 2016 ACM SIGCOMM Conference},
pages = {202–215},
numpages = {14},
location = {Florianopolis, Brazil},
series = {SIGCOMM '16}
}

@article{chen2022swing,
  title={Swing: Providing long-range lossless rdma via pfc-relay},
  author={Chen, Yanqing and Tian, Chen and Dong, Jiaqing and Feng, Song and Zhang, Xu and Liu, Chang and Yu, Peiwen and Xia, Nai and Dou, Wanchun and Chen, Guihai},
  journal={IEEE Transactions on Parallel and Distributed Systems},
  volume={34},
  number={1},
  pages={63--75},
  year={2022},
  publisher={IEEE}
}

@article{huang2024minimizing,
  title={Minimizing buffer utilization for lossless inter-DC links},
  author={Huang, Chengyuan and Xue, Feiyang and Yu, Peiwen and Wang, Xiaoliang and Chen, Yanqing and Wu, Tao and Han, Lei and Han, Zifa and Wang, Bingquan and Gong, Xiangyu and others},
  journal={IEEE/ACM Transactions on Networking},
  year={2024},
  publisher={IEEE}
}

@inproceedings{long2024lscc,
  title={Lscc: Link-segmented congestion control for rdma in cross-datacenter networks},
  author={Long, Minfei and Han, Jiangping and Wang, Wentao and Yang, Jiayu and Xue, Kaiping},
  booktitle={2024 IEEE/ACM 32nd International Symposium on Quality of Service (IWQoS)},
  pages={1--10},
  year={2024},
  organization={IEEE}
}

@article{zhang2021BDSInterDatacenter,
  title = {BDS+: An Inter-Datacenter Data Replication System With Dynamic Bandwidth Separation},
  shorttitle = {BDS+},
  author = {Zhang, Yuchao and Nie, Xiaohui and Jiang, Junchen and Wang, Wendong and Xu, Ke and Zhao, Youjian and Reed, Martin J. and Chen, Kai and Wang, Haiyang and Yao, Guang},
  year = {2021},
  month = apr,
  journal = {IEEE/ACM Transactions on Networking},
  volume = {29},
  number = {2},
  pages = {918--934},
  issn = {1558-2566},
}

@inproceedings{zhang2018BDSCentralized,
  title = {BDS: A Centralized near-Optimal Overlay Network for Inter-Datacenter Data Replication},
  shorttitle = {BDS},
  booktitle = {Proceedings of the Thirteenth EuroSys Conference},
  author = {Zhang, Yuchao and Jiang, Junchen and Xu, Ke and Nie, Xiaohui and Reed, Martin J. and Wang, Haiyang and Yao, Guang and Zhang, Miao and Chen, Kai},
  year = {2018},
  month = apr,
  pages = {1--14},
  publisher = {Association for Computing Machinery},
  address = {New York, NY, USA},
  isbn = {978-1-4503-5584-1}
}

@article{kandula2007DynamicLoad,
author = {Kandula, Srikanth and Katabi, Dina and Sinha, Shantanu and Berger, Arthur},
title = {Dynamic load balancing without packet reordering},
year = {2007},
issue_date = {April 2007},
publisher = {Association for Computing Machinery},
address = {New York, NY, USA},
volume = {37},
number = {2},
issn = {0146-4833},
journal = {SIGCOMM Comput. Commun. Rev.},
month = mar,
pages = {51–62},
numpages = {12}
}

\newpage
\appendix

%
\section{Artifact Appendix} 
\subsection{Abstract}
This artifact contains the complete implementation and evaluation framework for LCMP. The artifact includes: 

\begin{enumerate}[label=(\arabic*)]
    \item{An NS-3-based network simulator with implementations of ECMP, UCMP, and LCMP routing algorithms, see \texttt{simulation} folder.} 
    \item{Traffic generation tools supporting multiple realistic workload distributions (WebSearch, AliStorage, FbHdp), see \texttt{traffic\_gen} folder.}
    \item{Comprehensive analysis scripts for processing flow completion time (FCT) slowdown and link utilization metrics, see \texttt{analysis} folder.}
    \item{Automated batch experiment scripts to reproduce all figures and results presented in the paper, see \texttt{scripts} folder.}
\end{enumerate}

\subsection{Description \& Requirements}

\subsubsection{How to access}   
All code and data of the artifact is publicly available in the following GitHub repository:

\url{https://github.com/dyyuCS/LCMP}

A snapshot of this repository is archived at: 

\url{https://doi.org/10.5281/zenodo.18859753} 


The repository is licensed under the Apache-2.0 License.
For artifact evaluation, evaluators can access the repository directly to configure \& run our code locally.

\subsubsection{Hardware dependencies} 
Our full evaluation was conducted on a server with the following specifications:
\begin{itemize}
    \item CPU: 2 $\times$ AMD EPYC 7262 8-Core Processor (4+ cores recommended for parallel simulations) 
    \item RAM: 64 GB (8 GB minimum, 16 GB recommended for large-scale experiments)
    \item OS: Ubuntu 22.04 LTS
\end{itemize}

\textbf{Note:} The simulations are CPU-intensive. With the above configuration, experiments can be parallelized across multiple cores to significantly reduce wall-clock time. For systems with fewer cores, experiments will take proportionally longer but can still be run sequentially.

\subsubsection{Software dependencies}    
The artifact requires the following software dependencies:

\textbf{Core Dependencies:}
\begin{itemize}
    \item GCC/G++ 5.x (legacy compiler required for NS-3.17 compatibility)
    \item Python 2.7 (for NS-3 build system)
    \item Python 3.6+ (for traffic generation and analysis scripts)
    \item NS-3.18 network simulator (included in the repository)
    \item Mercurial, CMake, libboost-all-dev
    \item libsqlite3-dev, libxml2-dev, libgtk2.0-dev
\end{itemize}

\textbf{Python Packages (Python 3):}
\begin{itemize}
    \item numpy, pandas, matplotlib (for analysis and plotting)
    \item Standard library modules: argparse, subprocess, csv, os
\end{itemize}

\subsubsection{Benchmarks}

The artifact includes three realistic datacenter traffic workload distributions (WebSearch, AliStorage2019, FbHdp) used in our experiments. They are provided as CDF files in \texttt{traffic\_gen/flowCDF/} directory. The traffic generator (\texttt{traffic\_gen.py}) uses these distributions to generate synthetic inter-datacenter traffic at specified load levels, simulating realistic RDMA traffic patterns between geo-distributed datacenters.

\subsection{Set-up}

The artifact requires installation of system dependencies, building the NS-3 simulator, and installing Python packages for analysis scripts. Alternatively, evaluators can use the provided Docker-based environment to run simulations and analysis without manually installing all dependencies. Detailed step-by-step setup instructions are provided in the repository's main README file.

\subsection{Evaluation workflow}

\subsubsection{Major Claims}

The paper makes the following major claims about LCMP:

\begin{enumerate}[label=,
    itemindent=-1.7em, 
    leftmargin=1.7em,topsep=0pt]
    \item \circled{C1} \textbf{LCMP significantly reduces flow completion time (FCT) slowdown compared to ECMP and UCMP across different traffic loads.} This is demonstrated by experiments (E1, E2, E3) with results shown in \figref{fig:exp-testbed}, \figref{fig:exp-agg}, and \figref{fig:exp-dcpair} of the paper.
    
    \item \circled{C2} \textbf{LCMP effectively balances link utilization and reduces congestion in long-haul inter-datacenter links.} This is proven by experiment (E0) with results illustrated in \figref{fig:moti-total} of the paper.
    
    \item \circled{C3} \textbf{LCMP is robust across different traffic patterns and workloads.} This is validated by experiment (E4) with results shown in \figref{fig:exp-workloads} of the paper.
    
    \item \circled{C4} \textbf{LCMP works effectively with different RDMA congestion control algorithms.} This is demonstrated by experiment (E5) with results in \figref{fig:deep-cc} of the paper.
    
    \item \circled{C5} \textbf{Each component of LCMP's distributed cost function contributes to overall performance.} This is proven by experiment (E6) with ablation study results shown in \figref{fig:discuss-total} of the paper.
    
    \item \circled{C6} \textbf{LCMP maintains its performance advantages in large-scale deployments.} This is validated by experiments (E2, E3) with results in \figref{fig:exp-agg} and \figref{fig:exp-dcpair} of the paper.
\end{enumerate}

\subsubsection{Experiments}

This section provides information for reproducing all experiments presented in the paper. For convenience, we provide automated shell scripts in the \texttt{scripts/} folder that execute the complete workflow (simulation, analysis, and visualization) for each experiment. See \url{https://github.com/dyyuCS/LCMP/blob/main/scripts/README.md} for detailed usage instructions. Note that all figures in the paper were generated using Origin software based on the experimental data for better visual presentation.

\textbf{Experiment (E0): Link Utilization Analysis (Motivation)}

This experiment demonstrates the motivation for LCMP by showing how ECMP and UCMP create imbalanced link utilization in long-haul inter-datacenter links, while LCMP achieves better balance by considering both path characteristics and real-time congestion. This supports claim \circled{C2} and corresponds to \figref{fig:moti-total} in the paper. 

To run this experiment:

\noindent\fbox{\parbox{\dimexpr\linewidth-2\fboxsep-2\fboxrule}{%
\texttt{bash scripts/run\_figure1.sh}
}}

\vspace{5pt}

\textbf{Experiment (E1): Small-Scale Performance Comparison (8 DCs)}

This experiment compares LCMP, ECMP, and UCMP on an 8-datacenter topology across three traffic loads (30\%, 50\%, 80\%) using DCQCN. This supports claim \circled{C1} and corresponds to \figref{fig:exp-testbed} in the paper. 

To run this experiment:

\noindent\fbox{\parbox{\dimexpr\linewidth-2\fboxsep-2\fboxrule}{%
\texttt{bash scripts/run\_figure5.sh}
}}

\vspace{5pt}

\textbf{Experiment (E2): Large-Scale Performance Comparison (13 DCs)} 

This experiment evaluates LCMP scalability on a 13-datacenter geo-distributed topology across three traffic loads. This demonstrates LCMP's capability in large-scale inter-datacenter RDMA networks where centralized control is impractical. This supports claims \circled{C1} and \circled{C6}, and corresponds to \figref{fig:exp-agg} in the paper. 

To run this experiment:

\noindent\fbox{\parbox{\dimexpr\linewidth-2\fboxsep-2\fboxrule}{%
\texttt{bash scripts/run\_figure7\_8.sh}
}}

\vspace{5pt}

\textbf{Experiment (E3): Inter-DC Pair Analysis (13 DCs)}

This experiment analyzes performance between specific datacenter pairs (DC1-DC13) in the large-scale topology, focusing on long-haul paths with maximum geographic distance. This validates LCMP's effectiveness for the most challenging inter-datacenter scenarios. This supports claims \circled{C1} and \circled{C6}, and corresponds to \figref{fig:exp-dcpair} in the paper. 

To run this experiment:

\noindent\fbox{\parbox{\dimexpr\linewidth-2\fboxsep-2\fboxrule}{%
\texttt{bash scripts/run\_figure7\_8.sh}
}}

\vspace{5pt}

\textbf{Experiment (E4): Robustness Across Different Workloads}

This experiment evaluates LCMP with different traffic patterns (WebSearch, AliStorage, GoogleRPC). This supports claim \circled{C3} and corresponds to \figref{fig:exp-workloads} in the paper. 

To run this experiment:

\noindent\fbox{\parbox{\dimexpr\linewidth-2\fboxsep-2\fboxrule}{%
\texttt{bash scripts/run\_figure9.sh}
}}

\vspace{5pt}

\textbf{Experiment (E5): Robustness Across Different Congestion Control Algorithms}

This experiment tests LCMP with multiple RDMA transport protocols (DCQCN, HPCC, TIMELY, DCTCP). This demonstrates that LCMP's routing-layer improvements are orthogonal to and compatible with various RDMA congestion control mechanisms. This supports claim \circled{C4} and corresponds to \figref{fig:deep-cc} in the paper. 

To run this experiment:

\noindent\fbox{\parbox{\dimexpr\linewidth-2\fboxsep-2\fboxrule}{%
\texttt{bash scripts/run\_figure10.sh}
}}

\vspace{5pt}

\textbf{Experiment (E6): Ablation Study and Cost Function Analysis}

This experiment analyzes the contribution of each component in LCMP's distributed cost function, specifically examining how path costs and congestion costs work together for inter-datacenter routing decisions. This supports claim \circled{C5} and corresponds to \figref{fig:discuss-total} in the paper. 

To run this experiment:

\noindent\fbox{\parbox{\dimexpr\linewidth-2\fboxsep-2\fboxrule}{%
\texttt{bash scripts/run\_figure11\_ablation.sh}\\
\texttt{bash scripts/run\_figure11\_path\_cost.sh}\\
\texttt{bash scripts/run\_figure11\_congestion\_cost.sh}\\
\texttt{bash scripts/run\_figure11\_global\_weight.sh}
}}



\end{document}